# How to partition diversity

Richard Reeve[a,b,c,*], Tom Leinster[a,d], Christina A. Cobbold[a,e], Jill Thompson[a,f], Neil Brummitt[a,g], Sonia N. Mitchell[a,b], Louise Matthews[a,b]

*[a]Boyd Orr Centre for Population and Ecosystem Health and [b]Institute of Biodiversity, Animal Health and Comparative Medicine, College of Medical, Veterinary and Life Sciences, University of Glasgow, Glasgow, G12 8QQ, UK. [c]The Pirbright Institute, Ash Road, Pirbright, Woking, Surrey, GU24 0NF, UK. [d]School of Mathematics, College of Science and Engineering, University of Edinburgh, Edinburgh, EH9 3FD, UK. [e]School of Mathematics and Statistics, College of Science and Engineering, University of Glasgow, Glasgow, G12 8QW, UK. [f]Centre for Ecology & Hydrology, Bush Estate, Penicuik, Midlothian, EH26 0QB, UK. [g]Department of Life Sciences, Natural History Museum, London, SW7 5BD, UK.*

[*]Corresponding Author: Richard Reeve – richard.reeve@glasgow.ac.uk

# Abstract

Diversity measurement underpins the study of biological systems, but measures used vary across disciplines. Despite their common use and broad utility, no unified framework has emerged for measuring, comparing and partitioning diversity. The introduction of information theory into diversity measurement has laid the foundations, but the framework is incomplete without the ability to partition diversity, which is central to fundamental questions across the life sciences: How do we prioritise communities for conservation? How do we identify reservoirs and sources of pathogenic organisms? How do we measure ecological disturbance arising from climate change?

The lack of a common framework means that diversity measures from different fields have conflicting fundamental properties, allowing conclusions reached to depend on the measure chosen. This conflict is unnecessary and unhelpful. A mathematically consistent framework would transform disparate fields by delivering scientific insights in a common language. It would also allow the transfer of theoretical and practical developments between fields.

We meet this need, providing a versatile unified framework for partitioning biological diversity. It encompasses any kind of similarity between individuals, from functional to genetic, allowing comparisons between qualitatively different kinds of diversity. Where existing partitioning measures aggregate information across the whole population, our approach permits the direct comparison of subcommunities, allowing us to pinpoint distinct, diverse or representative subcommunities and investigate population substructure. The framework is provided as a ready-to-use R package to easily test our approach.

# Introduction

Diversity is a ubiquitous feature of biological systems, from the scale of the molecule to the rainforest. Its accurate assessment is fundamental to such contrasting problems as deciding which ecosystem should be prioritised for conservation, selecting the best sires for breeding, and determining the size of an epidemic. However, there has been great difficulty in agreeing on a common approach to quantifying and interpreting diversity. Much of the literature has converged on a framework for measuring diversity of single communities [1-3], which we explain below, but the unresolved issue remains that of determining how the diversity of subcommunities relates to that of the whole [4]. This *partitioning* of the diversity of a larger community among its constituent subcommunities is of fundamental importance in the life sciences, and is the goal of this paper.

**A need for diversity measures for meta and subcommunities** Throughout, we consider a *metacommunity* divided into *subcommunities*. We seek measures of alpha (within-subcommunity) diversity and beta (between-subcommunity) diversity, as well as gamma (total metacommunity) diversity, and we further seek to partition the metacommunity diversity into measurements for individual subcommunities.

Despite emerging agreement on how to measure diversity of a single community, whether taxonomic, functional, phylogenetic, etc., general approaches to measuring beta diversity are deficient [4-8]. A beta diversity measure should address two related sets of questions [9]: first, what is the effective number of distinct subcommunities present in a metacommunity [10]? And second, given a metacommunity divided into subcommunities, how distinct are the subcommunities, how evenly distributed are their *types* (*e.g.* species, sequences, phenotypes, or other biologically meaningful unit), what is their turnover [11], and which subcommunities are the least redundant? The second set of questions is concerned with how composition varies across a specific set of subcommunities such as along a spatial gradient or through a time series, whereas the first asks how many distinct subcommunities the metacommunity *naturally* divides into.

Numerous approaches to measuring beta diversity, as well as gamma and alpha diversity, have been proposed over many years [9,10,12,13]. In the simplest case, they assume that all subcommunities are distinct, as in Tuomisto's ${}^qD_\omega$[4]. Beyond that, they have focussed on satisfying either a simple multiplicative relationship between the components of diversity ($\alpha \times \beta = \gamma$) [4,5,10,14,15], or occasionally an additive relationship ($\alpha + \beta^+ = \gamma$) [14,16], where alpha is consistently defined as some kind of average [5] of subcommunity diversities. The advantages of multiplicative beta centre around the independence of alpha and beta diversity in determining gamma [17], although multiple definitions of independence often make it difficult to determine whether these requirements have been satisfied [8,14]. Tuomisto [4] reviewed many of these measures of beta diversity, but beyond agreement on the importance of some simple mathematical relationship between alpha, beta and gamma, no consensus has been reached on which is the best way to partition diversity.

**Identifying the underlying structure of a metacommunity** A unified framework for partitioning diversity should fulfill some fundamental and practical requirements – namely, the ability to identify sites of particular taxonomic, phylogenetic, phenotypic or functional interest *within* a metacommunity relative to its whole; either in terms of overall contribution to metacommunity diversity, or in terms of which subcommunities are the most diverse in their own right. To achieve this, we need to assign alpha, beta and gamma diversities to individual subcommunities, as well as to the metacommunity. This is a fundamental departure from the way that researchers have approached the partitioning question to date.

In this paper, we propose new subcommunity measures that satisfy these requirements. We then demonstrate that these measures can be combined to provide metacommunity level descriptors that are biologically important and able to describe: How similar in composition are the subcommunities to the metacommunity as a whole? Or, conversely, how many distinct subcommunities does the metacommunity naturally divide into (a kind of *beta diversity*)? How redundant are the subcommunities in terms of the resilience of the metacommunity to their loss? Or conversely, how concentrated are types in specific subcommunities; and what is their turnover (again, a kind of *beta diversity*)? We may also ask for the overall diversity of the metacommunity (*gamma diversity*) and the average diversity of its individual subcommunities (*alpha diversity*). However, we argue that it is from the subcommunity measures themselves that we extract the most useful information.

**Diversity of a single unpartitioned community** The recent agreement on the mathematical representation of the diversity of a single community [6] is rooted in the idea of the *effective* *number* of types. Using effective numbers means that the diversity of $S$ equally abundant types is $S$, which enables us to "speak naturally" [1] about diversity.

Jost [2] has argued convincingly that diversity measures should always be effective numbers, showing that such measures behave intuitively. However, it is also clear that measures of diversity should not just depend on the relative abundances of the different types being considered, but also on the similarities and differences between them. This is perhaps most evident in the assessment of functional diversity, when unrelated species can play similar functional roles in an ecosystem, but is also important for comparisons of diversity among species in the same genus versus those in different families, or for measuring diversity amongst individuals with differing amounts of shared phylogenetic history.

Addressing this challenge, Leinster and Cobbold [3], building on the work of Hill [1] and Ricotta and Szeidl [18], developed a unified similarity-sensitive diversity measure. Similarity between types is quantified in a *similarity matrix* $\mathbf{Z}$, where each entry $Z_{ii'}$ represents the similarity (usually between 0 and 1) between types $i$ and $i'$. This measure still has an effective number interpretation (that is, the effective number of distinct types), while elegantly encapsulating any notion of similarity or distance between types (whether functional, generating a measure of the effective number of distinct functional groups; taxonomic, the effective number of distinct taxonomic groups; phylogenetic, the effective number of distinct lineages; genetic, the effective number of distinct sequences; and so on). In the *naïve-type* case, where all types are viewed as completely distinct from one another, diversity is reduced to the original effective number formulation, the Hill numbers [1]. This approach is sufficiently general that it incorporates existing partial generalisations of diversity that include distance, such as that for phylogenetic similarity [19].

**Key properties of partitioning** Hill's effective number formulation of diversity [1] and Leinster and Cobbold's similarity-sensitive generalisation [3] are both based on Rényi's generalised entropies [20]. Given the power and generality of this entropy-based approach, a natural next step is to use Rényi's notion of generalised *relative* entropy to derive measures of the diversity of a subcommunity *relative* to the metacommunity as a whole. We leave the explanation of relative entropy and its role in partitioning diversity for Appendix S1.2, and focus instead on providing the reader with an intuitive understanding of these new formulae and their properties.

The measures we derive coincide with existing beta diversity measures [5] in degenerate cases (as proved in Appendix S2.1). Moreover, between them these measures satisfy all of the requirements we believe to be important for partitioning schemes. These requirements are mostly taken from the literature, include the formulation of relative diversities as effective numbers, and also include the important principles of replication (2, 25) and shattering (defined below).

Wilson and Shmida [17] propose that alpha and beta diversities should be independent, to ensure "useful application of a measure to systems with different alpha diversity". Jost [5] agrees, constraining their measurement by (a) requiring that "a given number should denote the same amount of diversity or uncertainty, whether it comes from the alpha component, the beta component, or the gamma component", and (b) generalising the definition of alpha diversity itself to "some type of average" of subcommunity diversities. He then clarifies independence as occurring when (c) "alpha is logically and mathematically unrelated to [beta]" [21]. These are Jost's first three properties of "intuitive alpha and beta", and we believe they form an excellent starting point for a family of diversity measures. They are all satisfied in our framework, as is his fifth requirement that alpha is never greater than gamma (see Appendix S2.1). However, we reject Jost's fourth requirement that gamma is entirely determined by alpha and beta; instead, we argue that the critical partitioning is not that of gamma into alpha and beta, but rather that of metacommunity alpha, beta and gamma into subcommunity alpha, beta and gamma.

We showcase our approach with examples from biodiversity and conservation, human demography, antimicrobial resistance and epidemiology, but first we present the theoretical framework.

# A subcommunity focus for alpha, beta and gamma diversity

We begin by introducing some core notation and key terminology, then explain each of the measures that appear in Tables 2 and 3. Following tradition, those measures are classified as "alpha", "beta" or "gamma" diversities. However, it will be useful to consider the reciprocal $\rho_j$ of the beta diversity $\beta_j$, and the reader is cautioned that we will refer to $\rho_j$ as a "beta diversity" too, despite the reciprocal relationship. Similar comments apply to the measures $\bar{\rho}_j$, $R$ and $\bar{R}$, defined below. We also discuss "normalised" and "raw" alpha and beta diversities (Tables 2 and 3, and below), where we account differently for the sizes of the subcommunities. Intuitively, we find that the normalised subcommunity measures tend to quantify diversity at the subcommunity level (the diversity of the subcommunity, the effective number of distinct subcommunities, etc.) and the raw subcommunity measures quantify it at the individual level (the average contribution to subcommunity or metacommunity diversity of each individual in the subcommunity, the redundancy across the metacommunity of each individual in the subcommunity, etc.). As a result, holding the metacommunity unchanged, changing the size of the subcommunity will change the raw subcommunity measures but not the normalised ones. In order to define our measures for a partitioned community we need to first review the measures for an unpartitioned community.

*Unpartitioned measures* At the core of effective number measures of diversity are the *Hill numbers* [1,22] (Table 1). These are ecological measures of species biodiversity derived from Rényi's generalisations [20] of Shannon entropy [23]. The Hill number of order $q$ is defined as a weighted power mean of order $1 - q$ that averages the inverses of the relative abundances of the $S$ types. We write $p_i$ for the relative abundance of type $i$ and weight the mean by $\boldsymbol{p} = (p_1, \ldots, p_S)$. The importance attached to relative abundance is reflected in the *viewpoint parameter,* $q$, which lies between 0 and ∞. Larger values of $q$ give increasingly *conservative* appraisals of diversity, by increasingly ignoring the rarer types. At the opposite extreme, the diversity at $q = 0$, which corresponds to species richness, is *anti-conservative*, attributing the same importance to rare species as to common ones by counting only the presence or absence of species. Famous instances of the Hill numbers are the effective number equivalents of Shannon entropy [23] and Simpson's concentration index [24] ($q = 1$ and $q = 2$, respectively). Leinster and Cobbold's [3] similarity-sensitive extension of the Hill numbers is defined in Table 1; a probabilistic interpretation is derived in Appendix S1.2.3.

**Table 1. Basic diversity formulae using effective numbers.**

| Name | Formula* |
|---|---|
| Power mean of order $r$ of $\boldsymbol{x}$ weighted by $\boldsymbol{u}$ | $\mathrm{M}_r(\boldsymbol{u},\boldsymbol{x}) = \begin{cases} \left[\sum_i u_i x_i^r\right]^{\frac{1}{r}} & r \neq 0^\dagger \\ \prod_i x_i^{u_i} & r = 0^\ddagger \end{cases}$ |
| Hill number | ${}^qD(\boldsymbol{p}) = \mathrm{M}_{1-q}(\boldsymbol{p}, 1/\boldsymbol{p})$ § |
| Similarity-sensitive diversity | ${}^qD^{\boldsymbol{Z}}(\boldsymbol{p}) = M_{1-q}(\boldsymbol{p}, 1/\boldsymbol{Zp})$ |

*$\boldsymbol{u} = (u_1, \ldots, u_n)$ and $\boldsymbol{x} = (x_1, \ldots, x_n)$ are vectors with $\sum_i u_i = 1$, and $r$ is a real number. The vector $\boldsymbol{p} = (p_1, \ldots, p_S)$ specifies the relative abundances $p_i$ of the S species in a community (so $p_i \geq 0$ and $\sum_i p_i = 1$), and $0 \leq q \leq \infty$ is the viewpoint parameter (see text). $\boldsymbol{Z} = (Z_{ii'})$ is an $n \times n$ matrix specifying similarities between the species (see text).* †*This formulation of the power means assumes that $u_i > 0$ for all i, otherwise such terms should be discounted.* ‡*At r = 0, 1/r is undefined, but the corresponding expression arises naturally in the limit.* §*Inverses of vectors are element-wise.*

*Notation for the partitioned measures* Now consider a metacommunity divided into $N$ subcommunities. The vector $\boldsymbol{p} = (p_1, \ldots, p_S)$ contains the relative abundances in the metacommunity. The abundance of type $i$ in subcommunity $j$ relative to the total metacommunity is $P_{ij}$, giving the matrix $\boldsymbol{P}$ of relative abundances, so that the *raw relative abundances* of the types in subcommunity $j$ are $\boldsymbol{P}_{.j} = (P_{1j}, \ldots, P_{Sj})$. Subcommunity $j$ constitutes a fraction $\sum_i P_{ij} = w_j$ of the total metacommunity abundance (so that $\sum_j w_j = 1$). It follows that $\boldsymbol{p} = \sum_j \boldsymbol{P}_{.j}$. We also denote the *normalised relative abundances* of the types in subcommunity $j$ in isolation – controlling for its size – as $\overline{\boldsymbol{P}_{.j}} = \boldsymbol{P}_{.j}/w_j$ (so that $\sum_i \overline{P_{ij}} = 1$). Correspondingly, we define both raw and normalised alpha and beta diversities. However, there is no distinction between normalised and raw gamma diversities, since we are only considering the case where $\sum_j w_j = \sum_i p_i = 1$, and we usually just refer to them as gamma diversity. Lastly, the similarity $Z_{ii'}$ between type $i$ and $i'$ is the $ii'$-entry of matrix $\boldsymbol{Z}$. Multiplying the matrix $\boldsymbol{Z}$ by the *column* vector $\boldsymbol{p}$ gives a vector $\boldsymbol{Zp}$ with entries $(\boldsymbol{Zp})_i = \sum_{i'} Z_{ii'} p_{i'}$, and similarly for $\boldsymbol{ZP}_{.j}$. As in the unpartitioned case, our partitioned diversity measures are defined as averages. We always use the term *average* to refer to a power mean of order $1 - q$ weighted by the relative sizes of the elements (Table 1, Power mean).

*Terminology*: To understand the measures we introduce, it is useful to keep in mind two extreme cases. The first is the *naïve-type* case, wherein different types have zero similarity ($\boldsymbol{Z} = \boldsymbol{I}$). The second is the *naïve-community* case (analysed fully in Appendix S2.2.4), wherein there are no shared types between subcommunities and types from different subcommunities have zero similarity (although similarities within a subcommunity may be non-zero).

**Table 2. Subcommunity diversity measures**

| **Formula** | **Description** |
|---|---|
| ${}^q\alpha_j^{\boldsymbol{Z}} = \mathrm{M}_{1-q}(\overline{\boldsymbol{P}}_{.j}, 1/\boldsymbol{ZP}_{.j})$ | **Raw alpha:** estimate of naïve-community* metacommunity diversity† |
| ${}^q\bar{\alpha}_j^{\boldsymbol{Z}} = \mathrm{M}_{1-q}(\overline{\boldsymbol{P}}_{.j}, 1/\boldsymbol{Z}\overline{\boldsymbol{P}}_{.j})$ | **Normalised alpha:** similarity-sensitive diversity of subcommunity *j* in isolation |
| ${}^q\rho_j^{\boldsymbol{Z}} = \mathrm{M}_{1-q}(\overline{\boldsymbol{P}}_{.j}, \boldsymbol{Zp}/\boldsymbol{ZP}_{.j})$ | **Raw beta (reversed):** redundancy of subcommunity *j* |
| ${}^q\beta_j^{\boldsymbol{Z}} = 1/{}^q\rho_j^{\boldsymbol{Z}}$ | **Raw beta:** distinctiveness of subcommunity *j* |
| ${}^q\bar{\rho}_j^{\boldsymbol{Z}} = \mathrm{M}_{1-q}(\overline{\boldsymbol{P}}_{.j}, \boldsymbol{Zp}/\boldsymbol{Z}\overline{\boldsymbol{P}}_{.j})$ | **Normalised beta (reversed):** representativeness of subcommunity *j* |
| ${}^q\bar{\beta}_j^{\boldsymbol{Z}} = 1/{}^q\bar{\rho}_j^{\boldsymbol{Z}}$ | **Normalised beta:** estimate of effective number of distinct subcommunities |
| ${}^q\gamma_j^{\boldsymbol{Z}} = \mathrm{M}_{1-q}(\overline{\boldsymbol{P}}_{.j}, 1/\boldsymbol{Zp})$ | **Gamma‡:** contribution per individual toward metacommunity diversity |

**Where different subcommunities are assumed to be distinct, i.e. there are no shared types between subcommunities and types from different subcommunities have zero similarity (although similarities*

*within a subcommunity may be non-zero).* [†]*Diversity as a descriptor in the body of this table refers to the effective number of types.* [‡]*Since we assume that* $\sum_i p_i = 1$, *we do not need to consider normalisation of* $\boldsymbol{w}, \boldsymbol{p}$, *or* $\gamma_j$.

**Table 3. Metacommunity diversity measures**

| Formula | Description |
|---|---|
| ${}^qA^{\boldsymbol{Z}} = \mathrm{M}_{1-q}(\boldsymbol{w}, {}^q\boldsymbol{\alpha}^{\boldsymbol{Z}})$[*] | **Raw alpha:** naïve-community[†] metacommunity diversity[‡] |
| ${}^q\bar{A}^{\boldsymbol{Z}} = \mathrm{M}_{1-q}(\boldsymbol{w}, {}^q\bar{\boldsymbol{\alpha}}^{\boldsymbol{Z}})$ | **Normalised alpha:** average diversity of subcommunities |
| ${}^qR^{\boldsymbol{Z}} = \mathrm{M}_{1-q}(\boldsymbol{w}, {}^q\boldsymbol{\rho}^{\boldsymbol{Z}})$ | **Raw beta (reversed):** average redundancy of subcommunities |
| ${}^qB^{\boldsymbol{Z}} = M_{1-q}(\boldsymbol{w}, {}^q\boldsymbol{\beta}^{\boldsymbol{Z}})$ | **Raw beta:** average distinctiveness of subcommunities |
| ${}^q\bar{R}^{\boldsymbol{Z}} = M_{1-q}(\boldsymbol{w}, {}^q\bar{\boldsymbol{\rho}}^{\boldsymbol{Z}})$ | **Normalised beta (reversed):** average representativeness of subcommunities |
| ${}^q\bar{B}^{\boldsymbol{Z}} = M_{1-q}(\boldsymbol{w}, {}^q\bar{\boldsymbol{\beta}}^{\boldsymbol{Z}})$ | **Normalised beta:** effective number of distinct subcommunities |
| ${}^qG^{\boldsymbol{Z}} = M_{1-q}(\boldsymbol{w}, {}^q\boldsymbol{\gamma}^{\boldsymbol{Z}})$ | **Gamma**[§]**:** metacommunity similarity-sensitive diversity |

[*]*Vectors are in bold, so* ${}^q\boldsymbol{\alpha}^{\boldsymbol{Z}} = ({}^q\alpha_1^{\boldsymbol{Z}}, \ldots, {}^q\alpha_N^{\boldsymbol{Z}})$ [†]*Where different subcommunities are assumed to be distinct, i.e. there are no shared types between subcommunities and types from different subcommunities have zero similarity (although similarities within a subcommunity may be non-zero).* [‡]*Diversity as a descriptor in the body of this table refers to the effective number of types.* [§]*Since we assume that* $\sum_i p_i = 1$, *we do not need to consider normalisation of* $\boldsymbol{w}, \boldsymbol{p}, \boldsymbol{\gamma}$ *or* $G$.

**Alpha diversities** Normalised subcommunity alpha diversity $\bar{\alpha}_j$ (Table 2) is a similarity-sensitive diversity, just as in Table 1, but of subcommunity $j$ in isolation. As with all of the subcommunity measures that follow, it does not depend on how the rest of the metacommunity is partitioned into subcommunities. Normalised metacommunity alpha diversity $\bar{A}$ (Table 3) is the average of the normalised subcommunity diversities, and is an effective number of distinct types. It lies between 1 and *S*, and reduces to Tuomisto's $\alpha_t$ in the naïve-type case [4].

Correspondingly, the raw subcommunity alpha diversity $\alpha_j$ (Table 2) is related to the normalised subcommunity alpha diversity via a rescaling by the size of the subcommunity ($\bar{\alpha}_j = w_j \times \alpha_j$), and measures diversity per individual in the subcommunity. In the absence of other knowledge, $\alpha_j$ can be used to estimate the metacommunity diversity, *G*. This estimate will be exact in the naïve-community case if every other $\alpha_j$ takes the same value, but in general will tend to be an overestimate due to commonalities between types in different subcommunities. Raw metacommunity alpha diversity *A* (Table 3) is the average of the raw subcommunity alpha diversities, and acts as an upper bound on the metacommunity gamma diversity *G* (Appendix S2.1.2). This allows us to constrain *G* without any knowledge of the relationships between or types within the subcommunities.

**Beta diversities** Four kinds of beta diversity naturally emerge from our framework, reflecting different aspects of the relationship between the metacommunity and its constituent subcommunities.

*Redundancy* The raw subcommunity beta diversity $\rho_j$ (Table 2) represents the redundancy of the subcommunity within the metacommunity, or the average redundancy of each individual in the subcommunity. It measures the extent to which the diversity of the metacommunity would be preserved if the subcommunity were to be destroyed. The redundancy $\rho_j$ is minimised (with value 1) when the types in subcommunity $j$ are minimally redundant: that is, if we lost the subcommunity then no individual in the remaining metacommunity would have any similarity to what was lost.

The redundancy $R$ of the metacommunity (Table 3) is the average of the $\rho_j$s, taking a minimum value of 1 in the naïve-community case (when there is no redundancy), and increasing as the subcommunities become more alike in their composition – again, be that through shared types or increased similarity between types. When all $N$ subcommunities in a metacommunity are identical in size and composition, then its redundancy, naturally, is $N$.

*Distinctiveness* Conversely, the reciprocal $\beta_j$ of $\rho_j$ (Table 2) measures the overall distinctiveness of a given subcommunity (or the concentration of types within it; see Appendix S1.2.3) or equivalently the distinctiveness of each individual within a given subcommunity relative to the metacommunity. It takes its maximum value of 1 when $\rho_j$ is minimised. This occurs when every individual in the subcommunity is completely dissimilar to every individual outside the subcommunity (Appendix S2.1.2). It is small when the subcommunity has much in common with the rest of the metacommunity, be that through shared types or high similarity between types. It can also be understood as a kind of turnover: not in the traditional sense, but between subcommunity $j$ *and the rest of the metacommunity*.

The average $B$ of the $\beta_j$s is a measure of the average distinctiveness of subcommunities (and lies between 0 and 1). To see the connection with turnover, consider a naïve-type case with each subcommunity containing the same number of types in equal abundance, and each type present in $k$ subcommunities, with a fraction 1/k changing from one subcommunity to the next ordered along a spatial gradient (see Appendix S2.2.7). Thus we have a turnover of $1/k$ along the gradient: and indeed, $\beta_j = 1/k$ (apart from at the ends), and $B \rightarrow 1/k$ as the number of subcommunities becomes large.

*Representativeness* The normalised subcommunity beta diversity $\bar{\rho}_j$ (Table 2) measures how representative, or typical, the subcommunity is of the metacommunity. It is simply a rescaling of $\rho_j$ (by a factor of $w_j$), and therefore behaves similarly. In particular, $\bar{\rho}_j$ is smallest when every individual in the subcommunity is completely dissimilar to every individual outside the subcommunity. In contrast to $\rho_j$, the minimum value of $\bar{\rho}_j$ is $w_j$. Consequently, at the minimum, a subcommunity is more representative when it constitutes more of the metacommunity. In the naïve-type case, the maximum value of $\bar{\rho}_j$ is 1, which is attained when the distribution of types in the subcommunity is identical to that of the metacommunity, meaning that the subcommunity represents the metacommunity faithfully. For general $\mathbf{Z}$, a low value of $\bar{\rho}_j$ indicates a subcommunity that has little in common with the metacommunity as a whole, and is in this sense not very representative of the metacommunity. It is often the case that similarities between individuals in the same subcommunity are, on average, greater than similarities between individuals in different subcommunities, and in this scenario $\bar{\rho}_j$ is less than 1 (an example in which $\bar{\rho}_j$ exceeds 1 can be found in Appendix S2.2.12). Taking the average representativeness of the subcommunities gives $\bar{R}$ (Table 3).

As an example, consider a naïve-type case with all types equally abundant in each subcommunity in which they are present, and all types equally abundant in the metacommunity as a whole. If every subcommunity contains only a proportion $r$ of the total number of types then $\bar{\rho}_j = \bar{R} = r$, reflecting the fact that each subcommunity represents a fraction $r$ of the total metacommunity (Appendix S2.2.8).

*Effective number of distinct subcommunities* In the absence of other knowledge, $\bar{\beta}_j = 1/\bar{\rho}_j$ can be used to estimate the effective number of subcommunities. To understand this, it is useful to consider $\bar{B}$, the average of the $\bar{\beta}_j$ s, which is the effective number of completely dissimilar subcommunities. Just as the effective number of distinct types is greatest when distinct types are equally abundant, the effective number of distinct subcommunities is greatest when distinct subcommunities are of equal size. Generally, $\bar{\beta}_j$ is an estimate of $\bar{B}$ based on subcommunity $j$, and is high when that subcommunity is both distinctive and small. For example, when the subcommunities are completely distinct (that is, in the naïve-community case), $\bar{B}$ is equal to the Hill number ${}^qD(\boldsymbol{w})$ (which Tuomisto [4] writes as ${}^qD_\omega$), and if all of the subcommunities are of equal size then $\bar{B}$ takes its maximum value of $N$.

**Gamma diversity** Finally, metacommunity gamma diversity, $G$ (Table 3), the diversity of the unpartitioned metacommunity (Table 1, Similarity-sensitive diversity), can also be expressed as an average of individual subcommunity gamma diversities, $\gamma_j$ (Table 2). See Appendix S1.1 for the proof of this equality. The quantity $\gamma_j$ measures the contribution per individual in the subcommunity to the diversity of the metacommunity as a whole, and reflects a different aspect of the relationship between the metacommunity and subcommunity $j$ than the beta diversity. It is a new kind of diversity measure, combining the inherent diversity of the subcommunity with its distinctiveness in the context of the metacommunity. The ability to identify sites that combine inherent (alpha) diversity with distinctiveness (beta) allows us to identify new patterns that are harder to observe using traditional measures of alpha and beta diversity.

Since $\gamma_j$ measures contribution to diversity per individual, an increase in $\bar{\alpha}_j$ does not necessarily cause a change in $\gamma_j$. Consider, for example, a naïve-type case with all types equally abundant in the metacommunity. If two subcommunities have evenly distributed types, but the first contains $k$ times as many types as the second, then $\bar{\alpha}_1 = k\bar{\alpha}_2$; but $\gamma_1 = \gamma_2$, because all the individuals concerned contribute equally to metacommunity diversity, being of types that are equally abundant in the metacommunity (see Appendix S2.2.9).

On the other hand, if two equally sized subcommunities have different constituent types, but with the same relative abundances, in such a way that all of the types in the first subcommunity are $k$ times rarer in the metacommunity than the types in the second, then $\bar{\alpha}_1 = \bar{\alpha}_2$ but $\gamma_1 = k\gamma_2$. This reflects the higher contribution of the first subcommunity to metacommunity diversity, through its rarer and therefore more distinct types (see Appendix S2.2.11).

**Overview** The subcommunity measures allow us to identify subcommunities with high inherent diversity, per individual or overall ($\alpha_j$ and $\bar{\alpha}_j$), high distinctiveness or that are very redundant in the metacommunity ($\beta_j$ or $\rho_j$), ones that are representative of large or only small, distinct parts of the metacommunity ($\bar{\rho}_j$ or $\bar{\beta}_j$), and ones with strong per-individual influence on metacommunity diversity ($\gamma_j$). Metacommunity gamma diversity ($G$) and normalised subcommunity alpha diversity ($\bar{\alpha}_j$) are simply our usual notions of diversity for an undivided group, while the other metacommunity and subcommunity diversity measures in Tables 2 and 3 are novel.

Our measures all depend on the parameter $q$. In particular, the exact interpretation of $\rho_j$ as the redundancy of subcommunity $j$ varies with $q$ (and similarly for $\beta_j$, $\bar{\beta}_j$ and $\bar{\rho}_j$). For instance, we can ask what it means for the redundancy of a subcommunity to achieve the minimal possible value, 1. When $q = 0$, anti-conservatively, this means that no type present in the subcommunity can be found anywhere else in the metacommunity, whereas at $q = \infty$, conservatively, it is sufficient for the subcommunity to have just one type not found anywhere else.

At $q = 1$, special relationships hold (connected to the well-established special properties of Shannon entropy); for instance, there are multiplicative relationships $\gamma_j = \alpha_j \times \beta_j = \bar{\alpha}_j \times \bar{\beta}_j$ and $G = A \times B = \bar{A} \times \bar{B}$. Similar expressions are found in the work of Routledge [22] and Jost [5], for example. Our metacommunity measures $\bar{A}$ and $\bar{B}$ play similar roles to Jost's alpha and beta diversities. In particular, $\bar{B}$ satisfies Jost's first three requirements for an intuitive beta diversity measure. Analogously to his fifth requirement that alpha is never greater than gamma, which for him is the equivalent of beta never being less than one, we have $\bar{B} \geq 1$ in the naïve-type case, which is the only case that he considers. See Appendix S2.1.2 for proofs.

Jost's alpha diversity (Tuomisto's $\alpha_R$) can be expressed as $A/{}^qD(\boldsymbol{w})$. In the naïve-community case, $A$ is precisely the metacommunity diversity $G$ and ${}^qD(\boldsymbol{w})$ is the beta diversity $\bar{B}$. Like our $\bar{A}$, Jost's alpha is a kind of average of the subcommunity diversities, but Jost's alpha weights the average diversity by the size of the subcommunities (with small subcommunities favoured for $q$ less than 1, and large subcommunities favoured for $q$ greater than 1), resulting in a measure that is not invariant under shattering (see below and Appendix S2.1.7). However, in the special case when the subcommunities are of equal size, $\bar{A}$ is exactly Jost's alpha.

**New properties of the new measures of diversity** New properties that we consider to be extremely important, are also enjoyed by these new diversity measures:

Normalised metacommunity alpha and beta diversities and gamma diversity are invariant under shattering of the subcommunities;

All subcommunity alpha, beta and gamma diversities are conditionally independent and therefore invariant to differences in partitioning of the rest of the metacommunity;

*All subcommunity alpha, beta and gamma diversities can therefore be directly compared* within a metacommunity to determine the relative merits of different subcommunities.

*Invariance under shattering* To explain what we mean by shattering, imagine a simple, idealised ecosystem with 4 genuine subcommunities that have at least some species in common (Fig 1a), for example subcommunities defined by geographical features or local abiotic conditions. Additionally, each subcommunity is assumed to be well-mixed, so if we geographically split a subcommunity (Fig 1b) the relative abundances of species in the new subcommunities will be the same as their parent subcommunity.

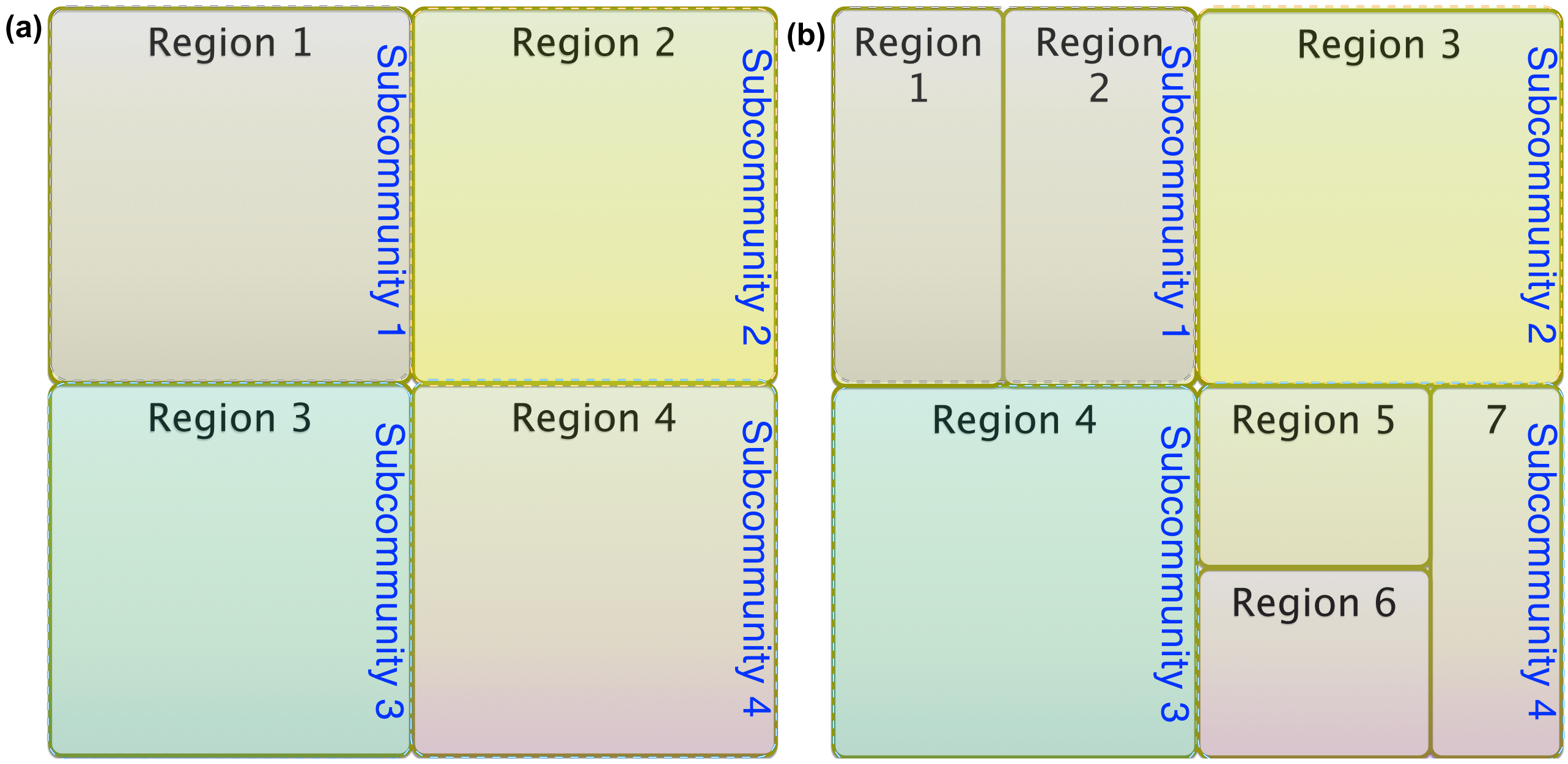

**Fig 1. An idealised ecosystem with four true subcommunities**. These have been identified (a) correctly – where each region (such as landscape or geographical area) reflects the true population subcommunity structure, and (b) incorrectly – as the subcommunities have been incorrectly further subdivided (*shattered*).

Suppose this ecosystem has a normalised beta diversity of $\bar{B} = b$ for some value of $q$. As the subcommunities have species in common and are therefore not completely distinct, $b < 4$ (since $b$ is an effective number). If, in our field study of the ecosystem, we correctly identify 4 geographically distinct subcommunities (Fig 1a, labelled as regions), then our diversity analysis should tell us that there are effectively $b$ distinct subcommunities. However, what happens if we incorrectly partition the ecosystem? Imagine that we think some geographical feature meaningfully divides two of the subcommunities, but that this turns out to be ecologically irrelevant (Fig 1b). Since drawing lines that do not cross subcommunity boundaries creates no genuinely new subcommunities, ***we claim that there must still be effectively b subcommunities.*** We refer to this as **invariance under shattering** – that breaking the subcommunities into smaller but identical parts must not affect beta diversity. Our new normalised metacommunity beta diversity measures ($\bar{R}$ and $\bar{B}$) satisfy the requirement that they are invariant under shattering (for proof, see Appendix S2.1.5), in contrast to Jost's simple multiplicative beta diversity [5] and most of the beta diversity measures listed by Tuomisto including the simplest naïve-community beta diversity, ${}^{q}D_{\omega}$ [4].

Our normalised metacommunity alpha diversity ($\bar{A}$) is also invariant under shattering. Some other metacommunity alpha diversities such as Tuomisto's $\alpha_t$ and "true alpha diversity" $\alpha_d$, which are actually both just $\bar{A}$ in the naïve-type case, are also invariant under shattering. In constrast, the manner of averaging suggested by Jost [5] (Tuomisto's $\alpha_R$), where metacommunity alpha diversity depends on $1/\sum w_j^q$, causes diversity to vary as subcommunities are broken up.

By definition gamma diversity does not depend on a specific decomposition into subcommunities. Therefore, any sensible definition of metacommunity gamma diversity (*e.g.* $G$) must also be invariant under shattering, since otherwise an arbitrary rearrangement of subcommunities might change gamma.

Our raw metacommunity alpha and beta diversity measures are not invariant under shattering, which is desirable, since $A$, the naïve-community estimate of metacommunity gamma diversity, should increase when smaller (shattered) communities are created with the same diversity; *B*, the average distinctiveness of the subcommunities, should decrease as the subcommunities are shattered into identical pieces; and *R*, the average redundancy of the subcommunities, should increase as more and more subcommunities are created with the same composition as each other.

*Conditional independence* Individual subcommunity diversities do not depend on the distributions or boundaries of other subcommunities. In order to be meaningful, the contributions of individual subcommunities to metacommunity alpha, beta and gamma diversities should be conditionally independent of each other – *i.e.* they should only depend on the composition of the other subcommunities through the composition of types across the metacommunity. Without this condition, the perceived importance of one subcommunity would be affected by arbitrary rearrangements of other subcommunity boundaries (despite these involving no gain or loss in individuals or types to the metacommunity as a whole), making any assessment of subcommunity diversity arbitrary. That conditional independence is satisfied is evident from inspection of the formulae for subcommunity diversities (Table 2), because they do not contain any terms for the relative abundances of the other subcommunities. The contribution from any given subcommunity to any metacommunity-level measure of alpha, beta and gamma diversity is therefore invariant with respect to the division of the rest of the metacommunity into subcommunities.

*Comparing subcommunities* Our new formulation provides measures of alpha, beta and gamma diversity at the subcommunity level that are conditionally independent of each other. Furthermore, they reflect their contribution to overall diversity in a manner that is invariant to changes in the partitioning of the rest of the metacommunity. We can therefore make direct, consistent comparisons between subcommunities that allow us to determine their relative merits in the context of the metacommunity as a whole.

# Conclusions

Diversity measurement underpins the study of biological systems across the life sciences. However, traditional measures have lacked consistency and comparability, both in their ability to account for different types of similarity between individuals, and by not providing a consistent means of partitioning diversity among subcommunities. Tackling these deficiencies, we have developed new measures of metacommunity alpha and beta diversity, that emerge naturally from information theory, and have shown that these can be used in any diversity context from taxonomic to functional, phylogenetic to phenotypic. A major advance of our formalism is the ability to decompose the metacommunity measures into their subcommunity contributions. These provide powerful tools with which to reveal biologically important subcommunities within a metacommunity, to assess the spatial and temporal variation in diversity, and to expose the metacommunity's true underlying subcommunity structure and dynamics.

From a mathematical perspective, we have shown that our new measures satisfy existing, desirable properties of diversity measures at the metacommunity level, but we reject the well-established [21] simple mathematical relationship between metacommunity alpha, beta and gamma, except in special cases. We have proposed new and biologically intuitive properties for diversity measures that allow subcommunities to be compared systematically and meaningfully. We have shown that our new measures satisfy these properties, while many existing measures do not. These key properties are:

**Invariance of normalized metacommunity measures under shattering**, in other words, the partitioning of subcommunities through biologically irrelevant divisions, does not affect measurements of normalised metacommunity alpha, beta or gamma diversity;

**Independence and invariance of subcommunity diversity measures** to differences in composition and partitioning of the rest of the metacommunity, indicating that the diversity of a subcommunity and its contribution to metacommunity diversity depend only on the composition of that subcommunity and of the metacommunity as a whole; and, arising from these properties:

**The ability to directly compare subcommunities** in terms of their intrinsic diversity (alpha), their redundancy (rho) or distinctiveness (beta) relative to the metacommunity as a whole, and their contribution to total metacommunity diversity (gamma).

By exposing subcommunity contributions, these diversity analyses offer a more refined and powerful tool than existing characterisations of "within", "between" and "total" diversity. Immediate applications of this framework extend well beyond biodiversity and conservation, with judicious choice of subcommunity structure (such as spatial and temporal) and similarity metric (whether functional or genetic) allowing us to use our new framework to tackle very diverse questions across the life sciences.

All of the functionality described above is provided as the ready-to-use R package "rdiversity" on github [25].

# Acknowledgements

We are grateful for the hospitality and support of the Centre de Recerca Matemàtica (CRM), Barcelona and the funding of BBSRC under grant BB/J020567/1 for supporting RR, LM, NB, CAC and TL to initiate this work, and BBSRC under grant number BB/P004202/1 for ongoing support of RR, TL, CAC and LM. TL was also funded by a grant from the Carnegie Trust. RR is also funded by BBSRC grants BB/L004070/1 and BB/L004828/1. LM is also funded by BBSRC grants BB/L004070/1, BB/F015313/1 and BB/K01126X/1 / NSF Award Number DEB1216040 and an EU-funded Marie Curie Initial Training Network (MC-ITN). SM is funded by BBSRC Doctoral Training Programme grant BB/J013854/1. Thanks also to Anne Chao and Lou Jost for valuable discussions that have improved this manuscript, and to CRM and BBSRC in supporting their attendance at the initial meetings at CRM. The BCI 50 ha plot was organized by S.P. Hubbell, R.B. Foster, R. Condit, S. Lao, and R. Perez under the Center for Tropical Forest Science and the Smithsonian Tropical Research in Panama. Numerous organizations have provided funding, principally the U.S. National Science Foundation, and hundreds of field workers have contributed.

# References

1. Hill MO. Diversity and Evenness: A Unifying Notation and Its Consequences. Ecology. 1973;54: 427–432. Available: http://www.jstor.org/stable/1934352

2. Jost L. Entropy and diversity. Oikos. 2006;113: 363–375. doi:10.1111/j.2006.0030-1299.14714.x

3. Leinster T, Cobbold CA. Measuring diversity: the importance of species similarity. Ecology. 2012;93: 477–489.

4. Tuomisto H. A diversity of beta diversities: straightening up a concept gone awry. Part 1. Defining beta diversity as a function of alpha and gamma diversity. Ecography. 2010;33: 2–22. doi:10.1111/j.1600-0587.2009.05880.x

5. Jost L. Partitioning diversity into independent alpha and beta components. Ecology. 2007;88: 2427–2439. Available: http://www.esajournals.org/doi/abs/10.1890/06-1736.1

6. Ellison AM. Partitioning diversity. Ecology. 2010;91: 1962–1963.

7. Tuomisto H. A diversity of beta diversities: straightening up a concept gone awry. Part 2. Quantifying beta diversity and related phenomena. Ecography. Blackwell Publishing Ltd; 2010;33: 23–45. doi:10.1111/j.1600-0587.2009.06148.x

8. Chao A, Chiu C-H, Hsieh TC. Proposing a resolution to debates on diversity partitioning. Ecology. 2012;93: 2037–2051.

9. Whittaker RH. Evolution and measurement of species diversity. Taxon. 1972;21: 213–251. doi:10.2307/1218190

10. Whittaker RH. Vegetation of the Siskiyou Mountains, Oregon and California. Ecol Monogr. 1960;30: 280–338.

11. Jaccard P. The Distribution of the Flora in the Alpine Zone. New Phytologist. New Phytologist Trust; 1912;11: 37–50.

12. Chiu C-H, Chao A. Distance-Based Functional Diversity Measures and Their Decomposition: A Framework Based on Hill Numbers. de Bello F, editor. PLoS ONE. Public Library of Science; 2014;9: e100014. doi:10.1371/journal.pone.0100014

13. Chiu C-H, Jost L, Chao A. Phylogenetic beta diversity, similarity, and differentiation measures based on Hill numbers. Ecol Monogr. 2014;84: 21–44. doi:10.1890/12-0960.1

14. Veech JA, Crist TO. Diversity partitioning without statistical independence of alpha and beta. Ecology. 2010;91: 1964–1969.

15. Baselga A. Multiplicative partition of true diversity yields independent alpha and beta components; additive partition does not. Ecology. Ecological Society of America; 2010;91: 1974–1981.

16. Veech JA, Summerville KS, Crist TO, Gering JC. The additive partitioning of species diversity: recent revival of an old idea. Oikos. 2002;99: 3–9. doi:10.1034/j.1600-0706.2002.990101.x

17. Wilson MV, Shmida A. Measuring Beta Diversity with Presence-Absence Data. J Ecol. British Ecological Society; 1984;72: 1055–1064.

18. Ricotta C, Szeidl L. Towards a unifying approach to diversity measures: bridging the gap between the Shannon entropy and Rao's quadratic index. Theor Popul Biology. 2006;70: 237–243. doi:10.1016/j.tpb.2006.06.003

19. Chao A, Chiu C-H, Jost L. Phylogenetic diversity measures based on Hill numbers. Philos T Roy Soc B. The Royal Society; 2010;365: 3599–3609. doi:10.1098/rstb.2010.0272

20. Rényi A. On measures of entropy and information. 1961. pp. 547–561.

21. Jost L. Independence of alpha and beta diversities. Ecology. 2010;91: 1969–1974.

22. Routledge RD. Diversity indices: which ones are admissible? J Theor Biol. 1979;76: 503–515.

23. Shannon CE. A Mathematical Theory of Communication. Bell Sys Tech J. Blackwell Publishing Ltd; 1948;27: 379–423. doi:10.1002/j.1538-7305.1948.tb01338.x

24. Simpson EH. Measurement of Diversity. Nature. 1949;163: 688–688.

25. Mitchell SN, Reeve R. rdiversity: A package for measuring similarity-sensitive diversity [Internet]. 2016. Available: https://github.com/boydorr/rdiversity

# S1 Appendix

## Derivations of subcommunity diversities and connections to entropy and information theory

The objective of this appendix is to present the derivation of the diversity partitioning measures presented in Table 2 of the main text and to elaborate on the connection between these measures and known quantities from information theory. Throughout this appendix we use the notation introduced in the main text. For ease of reading we present the explanations for a general $q \neq 1$ or $\infty$. However, by continuity of power means the results for the limiting cases of $q = 1$ and $q = \infty$ readily follow.

In section S1.1 we explain the concept of *uniqueness*, which is useful in interpreting the meaning of many of the formulas. Then in section S1.2 we show how our measures arise from a natural connection with entropy and information theory.

## S1.1 $G$, $\overline{\alpha}_j$ and $\gamma_j$ and the uniqueness of a type

Given a metacommunity with the relative abundances of the $S$ types denoted by the vector $\mathbf{p} = (p_1, \ldots, p_S)$ (with $0 \leq p_i \leq 1$ and $\sum_i p_i = 1$) and type similarity matrix $\mathbf{Z}$, where entry $Z_{ii'}$ denotes the similarity (from 0 to 1) between type $i$ and $i'$, we can calculate metacommunity gamma-diversity, $G$. Or put more simply we can calculate the similarity sensitive measure of the metacommunity's diversity (Leinster and Cobbold (2012)),

$$G = {}^qD^{\mathbf{Z}}(\mathbf{p}) = M_{1-q}\left(\mathbf{p}, \frac{1}{\mathbf{Zp}}\right) = \left(\sum_{i:p_i>0} p_i \left(\frac{1}{(\mathbf{Zp})_i}\right)^{1-q}\right)^{\frac{1}{1-q}} \tag{1}$$

($M_{1-q}\left(\mathbf{p}, \frac{1}{\mathbf{Zp}}\right)$ is a power mean, where $\mathbf{p}$ is the weight and $1/\mathbf{Zp}$ the quantity being averaged. See Table 1 of the main text for a definition of power means. Throughout the appendix we use the convention that the reciprocal of a vector refers to taking an element wise reciprocal.) An intuitive way to view the formulae for $G$ is to observe that

$$(\mathbf{Zp})_i = \sum_{i'=1}^{S} Z_{ii'} p_{i'} \tag{2}$$

is the expected similarity between an individual from the $i^{\text{th}}$ type and an individual chosen at random from the metacommunity. We can therefore think of

$$\frac{1}{(\mathbf{Zp})_i} = \frac{1}{\sum_{i'} Z_{ii'} p_{i'}} \tag{3}$$

as an indication of the *uniqueness* of the $i^{\text{th}}$ type. When type $i$ is not very similar to anything else in the metacommunity ($Z_{ii'} \approx 0$, for each $i'$) and if it is also rare ($p_i \ll 1$) then $\frac{1}{(\mathbf{Zp})_i}$ is large and that type can be viewed as being very unique. Taking a $1-q$ power mean of $\frac{1}{(\mathbf{Zp})_i}$, weighted by the relative abundances of type $i$ in the metacommunity, gives a measure of the average uniqueness of a type in the metacommunity and therefore a measure of diversity (Equation 1).

These are established ideas (see for example Leinster and Cobbold 2012) and the definition of metacommunity gamma-diversity ($G$) is largely uncontroversial. The definition of normalised subcommunity diversity, $\overline{\alpha}_j$, is also obtained in a straightforward way. If the metacommunity is partitioned into $N$ subcommunities then $\overline{\alpha}_j$ is the diversity of subcommunity $j$ when subcommunity $j$ is considered in isolation.

Consider an $S \times N$ matrix $\mathbf{P}$ of the relative abundances of the $S$ types present in the $N$ subcommunities. Then $P_{ij}$ is the abundance of type $i$ in subcommunity $j$ relative to the total abundance of the metacommunity as a whole. Hence $\mathbf{P}_{\cdot j} = (P_{1j}, \dots, P_{Sj})$ are the raw (may not sum to 1) relative abundances of types in subcommunity $j$, and $\overline{\mathbf{P}_{\cdot j}} = \mathbf{P}_{\cdot j}/w_j$ (where $w_j = \sum_i P_{ij}$ ) are the normalised (sum to 1) relative abundances of types in subcommunity $j$. The formulae for subcommunity diversity are then straightforward, and the natural relationship between the raw and normalised subcommunity diversities emerges,

$$\begin{aligned}\overline{\alpha}_j = {}^qD^Z(\overline{\mathbf{P}_{\cdot j}}) = M_{1-q}\left(\overline{\mathbf{P}_{\cdot j}}, \frac{1}{\mathbf{Z}\overline{\mathbf{P}_{\cdot j}}}\right) &= \left(\sum_{i:P_{ij}>0} \overline{\mathbf{P}_{ij}} \left(\frac{1}{\left(\mathbf{Z}\frac{P_{\cdot j}}{w_j}\right)_i}\right)^{1-q}\right)^{\frac{1}{1-q}} \\ = w_j M_{1-q}\left(\overline{\mathbf{P}_{\cdot j}}, \frac{1}{\mathbf{Z}\mathbf{P}_{\cdot j}}\right) = w_j \alpha_j.\end{aligned} \tag{4}$$

Continuing in this vein, we obtain a new quantity: $\gamma_j$, the subcommunity gamma-diversity. Subcommunity level gamma-diversity is the average uniqueness of a type from subcommunity $j$ as measured from the perspective of the metacommunity. This is given by taking the $1-q$ power mean of uniqueness of types in subcommunity $j$, but we measure uniqueness from the metacommunity's perspective (equation (3)). This is in contrast to $\overline{\alpha}_j$ where we measured uniqueness from the subcommunity's perspective. Thus

$$\gamma_j = M_{1-q}\left(\overline{\mathbf{P}_{\cdot j}}, \frac{1}{\mathbf{Zp}}\right) = \left(\sum_{i:P_{ij}>0} \frac{P_{ij}}{w_j} \left(\frac{1}{(\mathbf{Zp})_i}\right)^{1-q}\right)^{\frac{1}{1-q}}. \tag{5}$$

We can also refer to $\gamma_j$ as a *cross-diversity* which we elaborate on in Section S1.2.

An important observation is that the metacommunity gamma-diversity can be expressed as the $1-q$ power mean of the subcommunity gamma-diversities, weighted by the fraction of the metacommunity that constitutes the subcommunity,

$$M_{1-q}(\boldsymbol{w}, \boldsymbol{\gamma}) = \left(\sum_j w_j (\gamma_j)^{1-q}\right)^{\frac{1}{1-q}} = \left(\sum_j \sum_{i:P_{ij}>0} P_{ij} \left(\frac{1}{\mathbf{Zp}}\right)_i^{1-q}\right)^{\frac{1}{1-q}} = G \tag{6}$$

(here $\boldsymbol{\gamma} = (\gamma_1, \ldots, \gamma_N)$ and $\boldsymbol{w} = (w_1, \ldots, w_N)$ ). Equation (6) gives a decomposition of metacommunity gamma-diversity ($G$) into its subcommunity counterparts $\gamma_j$. We now have a very natural interpretation for $\gamma_j$ as the *contribution* of subcommunity $j$ to metacommunity diversity.

In the next section we will discuss the connection between the quantities we have discussed so far and notions of entropy from information theory. Our general approach borrows heavily from the theory of generalised entropy developed by Rényi (1961) and it is this connection that allows us to derive the remaining subcommunity diversity measures introduced in this paper.

## S1.2 Connections to entropy and information theory

Diversity indices have a longstanding relationship with 'entropies' of various kinds. In the case of $\mathbf{Z} = \mathbf{I}$, the diversities ${}^qD^{\mathbf{Z}}(\mathbf{p})$ are simple exponential transformations of the Rényi entropies ${}^qH(\mathbf{p})$,

$$ {}^qH(\mathbf{p}) = \ln\left({}^qD^{\mathbf{I}}(\mathbf{p})\right) = \frac{1}{1-q} \ln\left(\sum_{i:p_i>0} p_i^q\right) = \frac{1}{1-q} \ln\left(\sum_{i:p_i>0} p_i \left(\frac{1}{\mathbf{p}}\right)_i^{1-q}\right). \tag{7} $$

We can then define a similarity sensitive version of Rényi entropies, ${}^qH^{\mathbf{Z}}(\mathbf{p})$, where $\mathbf{Z}$ is no longer restricted to the identity matrix. Thus

$$ {}^qH^{\mathbf{Z}}(\mathbf{p}) = \ln\left({}^qD^{\mathbf{Z}}(\mathbf{p})\right) = \frac{1}{1-q} \ln\left(\sum_{i:p_i>0} p_i \left(\frac{1}{\mathbf{Zp}}\right)_i^{1-q}\right). \tag{8} $$

Ricotta and Szeidl (2006) proposed a similarity sensitive version of Tsallis entropy (a transformation of Rényi entropies), and Leinster and Cobbold (2012) defined and developed the similarity sensitive version ${}^qD^{\mathbf{Z}}(\mathbf{p})$ of the Rényi entropies. The indices ${}^qH^{\mathbf{Z}}(\mathbf{p})$ and ${}^qD^{\mathbf{Z}}(\mathbf{p})$ carry the same information, but only ${}^qD^{\mathbf{Z}}(\mathbf{p})$ has the virtue of being an effective number. So $G$ and $\overline{\alpha}_j$ have a direct connection to Rényi's entropies.

Patil and Taillie (1982) explained the connection between ecological diversity measures and information theory by relating information gain to our surprise at finding an individual of type $i$, denoted by the increasing function $\sigma\left(\frac{1}{(\mathbf{Zp})_i}\right)$. Our surprise increases with the uniqueness of type $i$, and so for a randomly chosen individual our expected surprise or information gain is $\sum p_i \sigma\left(\frac{1}{(\mathbf{Zp})_i}\right)$. We require $\sigma$ to have the property that the higher the probability of finding an individual of type $i$ the less information randomly selecting a type $i$ individual conveys. This and other reasonable properties in fact restricts the choice of $\sigma$ to precisely the function which leads to expected information gain being exactly Rényi's generalised entropies given in equation (7) (see Leinster and Cobbold (2012)).

### S1.2.1 Cross entropy and subcommunity gamma-diversity

We find that subcommunity gamma-diversity $\gamma_j$ is also related directly to entropy. Specifically, in the case of $\mathbf{Z} = \mathbf{I}$, $\gamma_j$ is the exponential of Rényi's cross-entropy (see Rao 2008 p31-33), hence the reason for referring to $\gamma_j$ as a cross-diversity earlier. If we consider two probability distributions given by $\mathbf{p}$ and $\mathbf{u}$, then we can define a similarity sensitive version of Rényi's cross-entropy given by

$$ {}^qH^{\mathbf{Z}}(\mathbf{p};\mathbf{u}) = \frac{1}{1-q} \ln \left( \sum_i p_i \left( \frac{1}{\mathbf{Zu}} \right)_i^{1-q} \right) \tag{9}$$

that describes the average information gained (in a similarity sensitive sense) about the probability distribution $\mathbf{p}$ by observing the probability distribution $\mathbf{u}$. The subcommunity gamma-diversity is

$$\gamma_j = M_{1-q} \left( \overline{\mathbf{P}_{\cdot j}}, \frac{1}{\mathbf{Zp}} \right) = \exp \left( {}^qH^{\mathbf{Z}} \left( \overline{\mathbf{P}_{\cdot j}}; \mathbf{p} \right) \right) \tag{10}$$

and can be thought of as the information gained about subcommunity $j$ by observing the metacommunity. In equation (10) we defined $\gamma_j$ as an average uniqueness of a species in subcommunity $j$ as viewed from the perspective of the metacommunity, and uniqueness is equivalent to information gain. Hence, the concepts of subcommunity gamma-diversity, cross entropy and information have natural connections. It should be pointed out that Rényi himself only defined entropies in the naïve case of $\mathbf{Z} = \mathbf{I}$ and so the cross entropy defined in equation (9) is a new generalisation of cross-entropy to the non-naïve case of $\mathbf{Z} \neq \mathbf{I}$.

### S1.2.2 Relative entropy and subcommunity beta-diversity

Rényi also defined generalised relative entropy (${}^qH(\mathbf{p}||\mathbf{u})$) which connects entropy and cross entropy to one another. In the simple case of $q = 1$ the relationship between these three quantities is

$$ {}^qH(\mathbf{p}||\mathbf{u}) = {}^qH(\mathbf{p};\mathbf{u}) - {}^qH(\mathbf{p}). \tag{11}$$

The relationship in equation (11) is more complicated for $q \neq 1$ (see Rao (2008) for a discussion of the relationship in the case of $q = 2$).

Our similarity sensitive generalisation of Rényi's relative entropy describes the divergence between the probability distribution $\mathbf{u}$ and the probability distribution $\mathbf{p}$ and is given by

$$ {}^qH^{\mathbf{Z}}(\mathbf{p}||\mathbf{u}) = \frac{1}{q-1} \ln \left( \sum_k p_k \left( \frac{\mathbf{Zp}}{\mathbf{Zu}} \right)_k^{q-1} \right) \tag{12}$$

(which in the case $\mathbf{Z} = \mathbf{I}$ reduces to ${}^qH(\mathbf{p}||\mathbf{u})$). Relative entropy is not a symmetrical quantity, so to be more precise, in the naïve case ($\mathbf{Z} = \mathbf{I}$) equation (12) gives a measure of the divergence of the probability distribution $\mathbf{p}$ *from* the probability distribution $\mathbf{u}$, and in the limiting case $q = 1$ reduces to the definition of Kullback-Leibler divergence.

Relative entropy fits quite naturally with notions of beta-diversity where one would want to measure the difference between the relative abundance of types in a subcommunity compared to in the metacommunity.

So we define the normalised subcommunity beta-diversity ($\bar{\beta}_j$) to be the exponential of relative entropy as follows,

$$\bar{\beta}_j = \exp\left({}^qH^{\mathbf{Z}}\left(\overline{\mathbf{P}_{\cdot j}}\middle\|\mathbf{p}\right)\right) = \frac{1}{M_{1-q}\left(\overline{\mathbf{P}_{\cdot j}}, \frac{\mathbf{Zp}}{\mathbf{Z}\overline{\mathbf{P}_{\cdot j}}}\right)} = \frac{1}{\left(\sum_{i:P_{ij}>0} \frac{P_{ij}}{w_j}\left(\frac{\mathbf{Zp}}{\mathbf{Z}\overline{\mathbf{P}_{\cdot j}}}\right)_i^{1-q}\right)^{\frac{1}{1-q}}}. \tag{13}$$

Returning to the concept of uniqueness we see that $\left(\frac{\mathbf{Zp}}{\mathbf{Z}\overline{\mathbf{P}_{\cdot j}}}\right)_i$ is the relative uniqueness of species $i$ in subcommunity $j$ (viewed as an isolated subcommunity) when compared to the uniqueness of type $i$ in the metacommunity. Normalised subcommunity beta-diversity ($\bar{\beta}_j$) is the reciprocal of the $1-q$ power mean of this relative uniqueness, weighted by the relative abundance of species in the subcommunity, when viewed in isolation. Hence, $\bar{\beta}_j$ would be high if the species in the subcommunity $j$ were on average not very unique/rare in the subcommunity, but very unique/rare in the metacommunity as a whole. We note that in the special case of $q = 1$ and $\mathbf{Z} = \mathbf{I}$, taking the exponentials of equation (11) gives the multiplicative relationship $\bar{\alpha}_j\bar{\beta}_j = \gamma_j$ discussed in the main text and that is advocated by Jost (2007) in his discussion of alpha, beta and gamma-diversity. In fact this is more generally true for $\mathbf{Z} \neq \mathbf{I}$, see Appendix S2.2.6 for details.

In addition, in the case $\mathbf{Z} = \mathbf{I}$ and $q = 1$, the metacommunity measures are also closely related to known entropy measures: $A$ is the exponential of the joint entropy of $\mathbf{p}$ and $\mathbf{w}$; $\bar{A}$ and $R$ are the exponentials of the conditional entropies of $\mathbf{p}$ given $\boldsymbol{w}$ and $\boldsymbol{w}$ given $\mathbf{p}$ (respectively); and $\bar{B}$ is the exponential of the mutual information of $\mathbf{p}$ and $\boldsymbol{w}$.

### S1.2.3 Understanding $\beta_j$ by sampling subcommunities and metacommunities

At this point it is useful to write down the raw beta-diversity for subcommunity $j$,

$$\beta_j = \frac{1}{M_{1-q}\left(\overline{\mathbf{P}_{\cdot j}}, \frac{\mathbf{Zp}}{\mathbf{Z}\overline{\mathbf{P}_{\cdot j}}}\right)} = \frac{w_j}{\left(\sum_{i:P_{ij}>0} \overline{P_{ij}}\left(\frac{\mathbf{Zp}}{\mathbf{Z}\overline{\mathbf{P}_{\cdot j}}}\right)_i^{1-q}\right)^{\frac{1}{1-q}}} = w_j\bar{\beta}_j. \tag{14}$$

Notice the difference in the power means that appear in

$$\bar{\beta}_j = \frac{1}{M_{1-q}\left(\overline{\mathbf{P}_{\cdot j}}, \frac{\mathbf{Zp}}{\mathbf{Z}\overline{\mathbf{P}_{\cdot j}}}\right)} \quad \text{and} \quad \beta_j = \frac{1}{M_{1-q}\left(\overline{\mathbf{P}_{\cdot j}}, \frac{\mathbf{Zp}}{\mathbf{Z}\overline{\mathbf{P}_{\cdot j}}}\right)} = w_j\bar{\beta}_j.$$

While $\beta_j$ gives a measure of the *distinctiveness* of subcommunity $j$, $\bar{\beta}_j$ scales this distinctiveness to account for the subcommunity's relative size within the metacommunity.

We can further understand $\beta_j$ and hence $\overline{\beta}_j$ by considering the special case of $\mathbf{Z} = \mathbf{I}$ and $q \in \mathbb{Z}$ with $q \geq 2$. Then we can interpret $\beta_j$ in terms of conditional probabilities. We have

$$\beta_j = \left( \sum_{i:P_{ij}>0} \frac{P_{ij}}{w_j} \left( \frac{P_{ij}}{p_i} \right)^{q-1} \right)^{\frac{1}{q-1}} \tag{15}$$

$$= \left( \sum_{i:\text{ type } i \text{ is present in subcommunity } j} \mathbb{P}(\text{type } i | \text{subcommunity } j) \mathbb{P}(\text{subcommunity } j | \text{type } i)^{q-1} \right)^{\frac{1}{q-1}}.$$

We can also estimate $\beta_j$ experimentally. Let us suppose we do the following experiment. We select an individual at random (with replacement) from subcommunity $j$; suppose this individual was of type $i$. Then we select at random $q-1$ individuals that are type $i$ from the metacommunity (with replacement). The event we are interested in is whether these $q-1$ individuals are *all* from subcommunity $j$. We record a one if the event has occurred and a zero otherwise. Let ${}^q\nu_j$ be the expected value of the outcome of this experiment, that is, the probability that the event described above has taken place. Then $\beta_j = ({}^q\nu_j)^{\frac{1}{q-1}}$.

We now see from equation (15) that $\beta_j$ increases as the probability of this event increases, so in other words if type $i$ were *concentrated* in subcommunity $j$ and not found in the rest of the metacommunity, and if this were the case for each type in subcommunity $j$, then $\nu_q$ would be very high, as would $\beta_j$.

A similar probabilistic interpretation can be given to $\beta_j$ for the case $q = 0$, $\mathbf{Z} = \mathbf{I}$:

$$\beta_j = \left( \sum_{i:P_{ij}>0} \frac{P_{ij}}{w_j} \left( \frac{P_{ij}}{p_i} \right)^{-1} \right)^{-1} = \left( \sum_{i:P_{ij}>0} \frac{p_i}{w_j} \right)^{-1} = \frac{w_j}{\sum_{i:P_{ij}>0} p_i} \tag{16}$$

$$= \frac{\text{number of individuals in subcommunity } j}{\text{number of individuals in the metacommunity that are of the types present in subcommunity } j}$$

$$= \mathbb{P}(\text{subcommunity } j | \text{type present in subcommunity } j)$$

The same expression can also be obtained directly from (15) by applying Bayes' theorem. So as with the case of $q \geq 2$, if the types present in subcommunity $j$ are not found anywhere else in the metacommunity then $\beta_j$ equals one, but if all the types are common in the rest of metacommunity then $\beta_j$ will approach $w_j$.

## References

1. Rao, S.M. (2008) Unsupervised learning: an information theoretic framework. Doctoral Thesis University of Florida.

2. Leinster, T. and Cobbold, C.A. (2012) Measuring diversity: the importance of species similarity. *Ecology*, 93, 477-489.

3. Rényi, A. (1961) On measures of entropy and information. In *Proc 4th Berkeley Symp Math Stat Prob 1960.* Presented at the 4th Berkeley Symposium on Mathematics, Statistics and Probability, 1960, pp 547-561.

4. Ricotta, C. and Szeidl, L. (2006) Towards a unifying approach to diversity measures: Bridging the gap between the Shannon entropy and Rao's quadratic index's. Theoretical Population Biology, 70, 237-243.

5. Patil, G.P. and Taillie, C. (1982) Diversity as a concept and its measurement. Journal of the American Statistical Association, 77, 548-567.

6. Jost, L. (2007) Partitioning diversity into independent alpha and beta components. *Ecology*, 88, 2427-2439.

# S2 Appendix

## Mathematical proofs and simple examples

Here we prove the mathematical assertions made in the main text, along with some further facts that clarify the meanings of our measures. We also illustrate their behaviour with some simple hypothetical examples.

**Conventions** We follow the notation of the main text. Unless indicated otherwise, the variable $i$ ranges over types $1, \ldots, S$ and the variable $j$ over subcommunities $1, \ldots, N$. The subcommunities are assumed not to be empty, so that $w_j > 0$ for all $j$.

In this appendix, we assume that the entries $Z_{ii'}$ of the $S \times S$ matrix $\boldsymbol{Z}$ satisfy $0 \leq Z_{ii'} \leq 1$ and $Z_{ii} = 1$. Most of the results only require $Z_{ii'} \geq 0$ and $Z_{ii} > 0$, but we make the stronger assumptions for the sake of simplicity.

We usually abbreviate ${}^q\alpha_j^{\boldsymbol{Z}}$, ${}^qA^{\boldsymbol{Z}}$, etc., as $\alpha_j$, $A$, etc. We write $\boldsymbol{\alpha} = (\alpha_1, \ldots, \alpha_N)$, and similarly for $\overline{\boldsymbol{\alpha}}$, $\boldsymbol{\beta}$, etc.

The vector $\boldsymbol{p}$ is a column vector, although for typesetting convenience we sometimes write $\boldsymbol{p} = (p_1, \ldots, p_S)$, as if it were a row vector. The same goes for $\boldsymbol{P}_{\cdot j}$, whenever $1 \leq j \leq N$.

For vectors $\boldsymbol{u} = (u_1, \ldots, u_n)$ and $\boldsymbol{v} = (v_1, \ldots, v_n)$, we write $\boldsymbol{u} \leq \boldsymbol{v}$ if $u_k \leq v_k$ for all $k \in \{1, \ldots, n\}$. We also write $\boldsymbol{u}/\boldsymbol{v}$ for the vector $(u_1/v_1, \ldots, u_n/v_n)$, and similarly $1/\boldsymbol{v} = (1/v_1, \ldots, 1/v_n)$. For instance, the diversities of Leinster and Cobbold [4] are given by ${}^qD^{\boldsymbol{Z}}(\boldsymbol{p}) = M_{1-q}(\boldsymbol{p}, 1/\boldsymbol{Zp})$, with the Hill numbers ${}^qD(\boldsymbol{p})$ as the special case $\boldsymbol{Z} = \boldsymbol{I}$.

**Power means** Given a probability vector $\boldsymbol{u} = (u_1, \ldots, u_n)$ (that is, a vector of nonnegative reals summing to 1) and another vector $\boldsymbol{x} = (x_1, \ldots, x_n)$ of nonnegative reals, the power means $M_r(\boldsymbol{u}, \boldsymbol{x})$ are defined as in Table 1 of the main text, for all real $r$. But it is also useful to define them for $r = \pm\infty$, which is done as follows:

$$M_{-\infty}(\boldsymbol{u}, \boldsymbol{x}) = \min_{i\colon u_i > 0} x_i, \qquad M_{\infty}(\boldsymbol{u}, \boldsymbol{x}) = \max_{i\colon u_i > 0} x_i.$$

These are the limits of $M_r(\boldsymbol{u}, \boldsymbol{x})$ as $r \to \pm\infty$, respectively.

We use some further facts about power means. They are increasing in their second arguments: if $\boldsymbol{x} \leq \boldsymbol{y}$ then $M_r(\boldsymbol{u}, \boldsymbol{x}) \leq M_r(\boldsymbol{u}, \boldsymbol{y})$ for all $r$ and $\boldsymbol{u}$. More substantially, $M_r$ is also increasing in $r$; that is, if $r \leq s$ then

$$M_r(\boldsymbol{u}, \boldsymbol{x}) \leq M_s(\boldsymbol{u}, \boldsymbol{x}) \tag{1}$$

for all $\boldsymbol{u}$ and $\boldsymbol{x}$. Furthermore, $M_r(\boldsymbol{u}, \boldsymbol{x})$ is continuous in $r$, so all of our subcommunity and metacommunity measures are continuous in $q$ (over the whole range $0 \leq q \leq \infty$).

All these facts are classical, and proofs can be found in texts such as [1].

## S2.1 Proofs

### S2.1.1 Simplified formulae for the metacommunity measures

In the main text, the metacommunity measures $A, \overline{A}, \ldots$ were defined as means of the subcommunity measures $\alpha_j, \overline{\alpha}_j, \ldots$, which are themselves defined as means. For a systematic approach, it is useful to define the **base terms**

$$a_{ij} = 1/(\boldsymbol{ZP}_{\cdot j})_i, \qquad \overline{a}_{ij} = 1/(\boldsymbol{Z}\overline{\boldsymbol{P}_{\cdot j}})_i,$$
$$r_{ij} = (\boldsymbol{Zp})_i/(\boldsymbol{ZP}_{\cdot j})_i, \qquad \overline{r}_{ij} = (\boldsymbol{Zp})_i/(\boldsymbol{Z}\overline{\boldsymbol{P}_{\cdot j}})_i,$$
$$g_{ij} = 1/(\boldsymbol{Zp})_i.$$

The definitions of the subcommunity measures can be expressed uniformly using the base terms: writing $\boldsymbol{a}_{\cdot j} = (a_{1j}, \ldots, a_{Sj})$ etc., we have

$$\alpha_j = M_{1-q}(\overline{\boldsymbol{P}_{\cdot j}}, \boldsymbol{a}_{\cdot j}), \quad \overline{a}_j = M_{1-q}(\overline{\boldsymbol{P}_{\cdot j}}, \overline{\boldsymbol{a}}_{\cdot j}),$$
$$\rho_j = M_{1-q}(\overline{\boldsymbol{P}_{\cdot j}}, \boldsymbol{r}_{\cdot j}), \quad \overline{\rho}_j = M_{1-q}(\overline{\boldsymbol{P}_{\cdot j}}, \overline{\boldsymbol{r}}_{\cdot j}), \quad \beta_j = 1/\rho_j, \quad \overline{\beta}_j = 1/\overline{\rho}_j,$$
$$\gamma_j = M_{1-q}(\overline{\boldsymbol{P}_{\cdot j}}, \boldsymbol{g}_{\cdot j}).$$

The metacommunity measures are *defined* as means of the subcommunity measures, but all of them except $B$ and $\overline{B}$ can also be expressed directly as means of the base terms:

$$A = M_{1-q}(\boldsymbol{P}, \boldsymbol{a}), \qquad \overline{A} = M_{1-q}(\boldsymbol{P}, \overline{\boldsymbol{a}}),$$
$$R = M_{1-q}(\boldsymbol{P}, \boldsymbol{r}), \qquad \overline{R} = M_{1-q}(\boldsymbol{P}, \overline{\boldsymbol{r}}),$$
$$G = M_{1-q}(\boldsymbol{P}, \boldsymbol{g}) = M_{1-q}(\boldsymbol{p}, 1/\boldsymbol{Zp}) = {}^qD^{\boldsymbol{Z}}(\boldsymbol{p}).$$

This means, for instance, that $A$ is the power mean of order $1-q$ of the quantities $a_{ij}$, weighted by $P_{ij}$. So for $q \neq 1, \infty$,

$$A = \Bigg( \sum_{i,j\colon P_{ij}>0} P_{ij} a_{ij}^{1-q} \Bigg)^{1/(1-q)}.$$

The simplified formulae for the metacommunity measures follow routinely from the definitions. No such simplification occurs for $B$ and $\overline{B}$.

### S2.1.2 Upper and lower bounds

Here we establish the ranges of possible values taken by our measures, and describe cases in which they achieve their extremal values. The results are summarised in Table S2.1. The extremal cases are easily verified, and the inequalities are proved below.

We will use the fact that the measures ${}^qD^{\boldsymbol{Z}}$ of Leinster and Cobbold [4] always take values between 1 and $S$, for a community of $S$ types. In particular, this is true of the Hill numbers ${}^qD = {}^qD^{\boldsymbol{I}}$. We write $=^*$ or $\leq^*$ for an equality or inequality that holds in the naïve-type case $\boldsymbol{Z} = \boldsymbol{I}$, but not necessarily for all $\boldsymbol{Z}$.

**Alpha-diversities** By the observations just made, $1 \leq \overline{\alpha}_j \leq S$ for all $j$, so $1 \leq \overline{A} \leq S$. This reflects the fact that $\overline{A}$ is an effective number of types.

It follows that $1/w_j \leq \alpha_j \leq S/w_j$ for all $j$. Hence

$$M_{1-q}(\boldsymbol{w}, 1/\boldsymbol{w}) \leq A \leq M_{1-q}(\boldsymbol{w}, S/\boldsymbol{w}),$$

or equivalently

$${}^qD(\boldsymbol{w}) \leq A \leq {}^qD(\boldsymbol{w}) \cdot S.$$

But $1 \leq {}^qD(\boldsymbol{w}) \leq N$, so $1 \leq A \leq NS$.

**Beta-diversities** First, fix $j$. We have $\boldsymbol{p} = \sum_{j'} \boldsymbol{P}_{\cdot j'} \geq \boldsymbol{P}_{\cdot j}$, and so

$$\boldsymbol{Zp} \geq \boldsymbol{ZP}_{\cdot j}. \tag{2}$$

Hence

$$\rho_j = M_{1-q}\bigg( \overline{\boldsymbol{P}_{\cdot j}}, \frac{\boldsymbol{Zp}}{\boldsymbol{ZP}_{\cdot j}} \bigg) \geq 1.$$

| (a) | Subcommunity | Metacommunity |
| --- | --- | --- |
| Raw alpha | $1/w_j \leq \alpha_j \leq S/w_j$ | $1 \leq A \leq NS$ |
| Normalised alpha | $1 \leq \overline{\alpha}_j \leq S$ | $1 \leq \overline{A} \leq S$ |
| Raw beta | $1 \leq \rho_j \leq^* 1/w_j$<br>$w_j \leq^* \beta_j \leq 1$ | $1 \leq R \leq^* N$<br>$0 < B \leq 1$ |
| Normalised beta | $w_j \leq \overline{\rho}_j \leq^* 1$<br>$1 \leq^* \overline{\beta}_j \leq 1/w_j$ | $0 < \overline{R} \leq^* 1$<br>$1 \leq^* \overline{B} \leq N$ |
| Gamma | $1 \leq \gamma_j \leq S/w_j$ | $1 \leq G \leq S$ |
| Alpha vs. gamma | $\gamma_j \leq \alpha_j$ | $G \leq A \leq^* NG$<br>$\overline{A} \leq^* G$ |

| (b) | Subcommunity | Metacommunity |
| --- | --- | --- |
| Alpha | $\alpha_j = 1/w_j$, $\overline{\alpha}_j = 1$:<br>subcommunity $j$ is single-type<br>$\alpha_j = S/w_j$, $\overline{\alpha}_j = S$:<br>naïve-type & subcomm $j$ balanced | $A = 1$: only one subcomm & only one type<br>$\overline{A} = 1$: every subcommunity is single-type<br>$A = NS$: naïve-type, each subcomm balanced, & subcommunities equal size<br>$\overline{A} = S$: naïve-type & each subcomm balanced |
| Beta | $\rho_j = 1$, $\beta_j = 1$, $\overline{\rho}_j = w_j$, $\overline{\beta}_j = 1/w_j$:<br>subcommunity $j$ is isolated<br>$\rho_j =^* 1/w_j$, $\beta_j =^* w_j$, $\overline{\rho}_j =^* 1$, $\overline{\beta}_j =^* 1$:<br>distribution of types in subcomm $j$ is same as in metacommunity | $R = 1$, $B = 1$: naïve-community<br>$\overline{B} = N$: naïve-comm & subcomms equal size<br>$R =^* N$: well-mixed & subcomms equal size<br>$\overline{R} =^* 1$, $\overline{B} =^* 1$: well-mixed |
| Gamma | $\gamma_j = 1$: every type in subcomm $j$ is identical to every type in metacomm<br>$\gamma_j = S/w_j$: naïve-type, types balanced, & subcomm $j$ is whole metacomm | $G = 1$: all types in metacomm are identical<br>$G = S$: naïve-type & types are balanced |
| Gamma vs. alpha | $\gamma_j = \alpha_j$: subcommunity $j$ is isolated | $G = A$: naïve-community<br>$A =^* NG$: well-mixed & subcomms equal size<br>$\overline{A} =^* G$: well-mixed |

Table S2.1: (a) The ranges of our measures and inequalities between them. The symbol $\leq^*$ denotes an inequality that holds in the naïve-type model $\boldsymbol{Z} = \boldsymbol{I}$, but not for all $\boldsymbol{Z}$. (See Section S2.2.12 for counterexamples.) Some of the metacommunity measures also satisfy sharper inequalities, involving $\boldsymbol{w}$; see the text of Section S2.1.2. (b) Extremal values. For example, the first entry states that if subcommunity $j$ contains only one type then $\alpha_j$ and $\overline{\alpha}_j$ attain their minimum values. The symbol $=^*$ denotes an equality that holds in the naïve-type model $\boldsymbol{Z} = \boldsymbol{I}$, but not necessarily for all $\boldsymbol{Z}$. **Balanced** means that all types have equal abundance. **Well-mixed** means that all subcommunities have the same distribution of types ($\overline{\boldsymbol{P}_{\cdot j}} = \boldsymbol{p}$ for all $j$). A subcommunity is **isolated** if the types present in it are different from, and completely dissimilar to, all the types present in other subcommunities. **Naïve-community** means that every subcommunity is isolated.

In the naïve-type case $\boldsymbol{Z} = \boldsymbol{I}$, inequality (1) gives

$$\rho_j \leq^* M_1\left(\overline{\boldsymbol{P}_{\cdot j}}, \frac{\boldsymbol{p}}{\boldsymbol{P}_{\cdot j}}\right) = \sum_{i\colon P_{ij}>0} \frac{p_i}{w_j} \leq \sum_{i=1}^{S} \frac{p_i}{w_j} = \frac{1}{w_j}. \tag{3}$$

Thus, $1 \leq \rho_j \leq^* 1/w_j$. The inequalities for $\beta_j$, $\overline{\rho}_j$ and $\overline{\beta}_j$ in Table S2.1(a) follow immediately.

Now we prove the inequalities for the metacommunity beta diversities. The inequalities for $\overline{\rho}_j$ and $\beta_j$ immediately imply that

$$\min_j w_j \leq \overline{R} \leq^* 1, \qquad \min_j w_j \leq^* B \leq 1,$$

giving $0 < \overline{R} \leq^* 1$ and $0 < B \leq 1$. It is also immediate that $R \geq 1$ and $\overline{B} \geq^* 1$. It remains to prove that $R \leq^* N$ and $\overline{B} \leq N$.

To prove that $R \leq^* N$, recall from (3) that $\rho_j \leq^* 1/w_j$ for all $j$; so

$$R = M_{1-q}(\boldsymbol{w}, \boldsymbol{\rho}) \leq^* M_{1-q}(\boldsymbol{w}, 1/\boldsymbol{w}) = {}^qD(\boldsymbol{w}).$$

Hence $R \leq^* {}^qD(\boldsymbol{w})$, and in particular, $R \leq^* N$.

Similarly, since $\overline{\beta}_j \leq 1/w_j$ for all $j$, we have $\overline{B} \leq {}^qD(\boldsymbol{w})$ and so $\overline{B} \leq N$. That $\overline{B} \leq N$ reflects the fact that $\overline{B}$ is an effective number of subcommunities.

**Gamma-diversities** We have $(\boldsymbol{Zp})_i = \sum_{i'} Z_{ii'} p_{i'}$, so that $(\boldsymbol{Zp})_i$ is a mean of entries of $\boldsymbol{Z}$, all of which are less than or equal to 1; hence $(\boldsymbol{Zp})_i \leq 1$ for all $i$. So $g_{ij} = 1/(\boldsymbol{Zp})_i \geq 1$ for all $i, j$, giving $\gamma_j \geq 1$ for all $j$.

On the other hand, we are about to prove that $\gamma_j \leq \alpha_j$, and we proved earlier that $\alpha_j \leq S/w_j$; hence $\gamma_j \leq S/w_j$.

Finally, $G = {}^qD^{\boldsymbol{Z}}(\boldsymbol{p})$, so $1 \leq G \leq S$. This reflects the fact that the metacommunity diversity $G$ is an effective number of types.

**Comparison between raw alpha- and gamma-diversities** Fix $j$. We have $g_{ij} \leq a_{ij}$ for all $i$ by (2), so

$$\gamma_j = M_{1-q}(\overline{\boldsymbol{P}_{\cdot j}}, \boldsymbol{g}_{\cdot j}) \leq M_{1-q}(\overline{\boldsymbol{P}_{\cdot j}}, \boldsymbol{a}_{\cdot j}) = \alpha_j.$$

This holds for all $j$, so $G \leq A$. Equality is attained in the naïve-community model (Section S2.2.4).

We now show that $A \leq^* NG$. Put $\boldsymbol{Z} = \boldsymbol{I}$. By the formula for $A$ in Section S2.1.1,

$$A =^* M_{1-q}(\boldsymbol{P}, 1/\boldsymbol{P}) = {}^qD(\boldsymbol{P}).$$

Now for each $i$ such that $p_i > 0$, there is a probability vector $\boldsymbol{P}_{i\cdot}/p_i$. This vector is $N$-dimensional, so ${}^qD(\boldsymbol{P}_{i\cdot}/p_i) \leq N$. For $q < 1$, this gives

$$\sum_{j\colon P_{ij}>0} P_{ij}^q \leq N^{1-q} p_i^q,$$

and for $1 < q < \infty$, the inequality is reversed. Summing both sides over $i$ and raising to the power of $1/(1-q)$ gives, in both cases,

$$A =^* \left( \sum_{i,j\colon P_{ij}>0} P_{ij}^q \right)^{1/(1-q)} \leq N \left( \sum_i p_i^q \right)^{1/(1-q)} = NG.$$

Hence $A \leq^* NG$ for all $q \neq 1, \infty$. But $A$ and $G$ are continuous in $q$, so $A \leq^* NG$ for all $1 \leq q \leq \infty$.

(Given the sharp forms of the inequalities for the other metacommunity measures, it is tempting to conjecture that $A \leq^* {}^qD(\boldsymbol{w})G$. However, this is in general false.)

Thus, $A/N \leq^* G \leq A$, or equivalently $1 \leq A/G \leq^* N$. One can interpret $A/G$ as a generalised Jaccard index. Indeed, consider the case $\boldsymbol{Z} = \boldsymbol{I}$, $N = 2$ and $q = 0$. Then our inequalities become $0 \leq A/G - 1 \leq 1$, and in fact, $A/G - 1$ is precisely the Jaccard index $a/(a+b+c)$. Here $a$ is the number of types present in both subcommunities, $b$ is the number present in the first only, and $c$ is the number present in the second only.

**Comparison between normalised alpha- and gamma-diversities** We show that $\overline{A} \leq^* G$. Since both $\overline{A}$ and $G$ are continuous in $q$, it suffices to prove this for $q \neq 0, 1, \infty$.

Put $\boldsymbol{Z} = \boldsymbol{I}$. Then by the formula for $\overline{A}$ in Section S2.1.1,

$$\overline{A} =^* \left( \sum_{i,j\colon P_{ij}>0} P_{ij} \overline{P_{ij}}^{q-1} \right)^{1/(1-q)} = \left( \sum_{i,j} w_j^{1-q} P_{ij}^q \right)^{1/(1-q)},$$

and $G =^* \left( \sum_i p_i^q \right)^{1/(1-q)}$, so we must prove that

$$\left( \sum_{i,j} w_j^{1-q} P_{ij}^q \right)^{1/(1-q)} \leq \left( \sum_i p_i^q \right)^{1/(1-q)}. \tag{4}$$

We split the proof into two cases: $0 < q < 1$ and $1 < q < \infty$. Both will use a version of Hölder's inequality (Theorem 13 of [1]), which states that whenever $s$ and $t$ are positive real numbers satisfying $1/s + 1/t = 1$,

$$\sum_j x_j y_j \leq \left( \sum_j x_j^s \right)^{1/s} \cdot \left( \sum_j y_j^t \right)^{1/t}$$

for all $x_1, \ldots, x_N, y_1, \ldots, y_N \geq 0$.

In the case $0 < q < 1$, we apply Hölder's inequality with $s = 1/(1-q)$, $t = 1/q$, $x_j = w_j^{1-q}$ and $y_j = P_{ij}^q$. This gives

$$\sum_j w_j^{1-q} P_{ij}^q \leq \left( \sum_j w_j \right)^{1-q} \cdot \left( \sum_j P_{ij} \right)^q = p_i^q$$

for each $i$. Summing this inequality over all $i$ and raising both sides to the power of $1/(1-q)$ gives (4).

In the case $1 < q < \infty$, we apply Hölder's inequality with $s = q/(q-1)$, $t = q$, $x_j = w_j^{(q-1)/q}$ and $y_j = w_j^{(1-q)/q} P_{ij}$. This gives

$$\sum_j P_{ij} \leq \Big(\sum_j w_j\Big)^{(q-1)/q} \cdot \Big(\sum_j w_j^{1-q} P_{ij}^q\Big)^{1/q}$$

or equivalently

$$p_i^q \leq \sum_j w_j^{1-q} P_{ij}^q.$$

Summing this inequality over all $i$ and raising both sides to the power of $1/(q-1)$ (which is negative) gives (4), completing the proof.

**Monotonicity in $q$** Since the power means $M_r$ are increasing in $r$, each of $\alpha_j$, $\overline{\alpha}_j$, $\rho_j$, $\overline{\rho}_j$ and $\gamma_j$ is decreasing in $q$. For the same reason, using the formulae for the metacommunity measures in Section S2.1.1, each of $A$, $\overline{A}$, $R$, $\overline{R}$ and $G$ is also decreasing in $q$.

Moreover, since $\rho_j$ and $\overline{\rho}_j$ are decreasing in $q$, both $\beta_j$ and $\overline{\beta}_j$ are increasing in $q$. But $B$ and $\overline{B}$ are in general neither increasing nor decreasing in $q$. For example, take $\boldsymbol{P} = \left(\begin{smallmatrix} 0.1 & 0.2 \\ 0.6 & 0.1 \end{smallmatrix}\right)$ and $\boldsymbol{Z} = \boldsymbol{I}$; then plotting $B$ and $\overline{B}$ against $q$ shows that neither is monotone.

### S2.1.3 Replication

Consider a large region—a 'multicommunity'—divided into $m$ equal-sized metacommunities, each of which is further divided into $N$ subcommunities. Thus, the multicommunity is divided into $mN$ subcommunities. Suppose that the $m$ metacommunities are identical in every way (the same subcommunity divisions, the same similarities between types, and the same relative abundances) except that each uses a fresh set of types, and types in different metacommunities are completely dissimilar. Proposition A11 of Leinster and Cobbold [4] shows that the diversity of the multicommunity is $m$ times the diversity of any one of its constituent metacommunities. What about the various alpha- and beta-diversities?

Let us use starred notation ($\boldsymbol{P}^\star$, $\boldsymbol{Z}^\star$, $\alpha_j^\star$, ...) for the matrices and diversity measures associated with the multicommunity divided into subcommunities (ignoring the metacommunity level), and, as usual, unstarred notation ($\boldsymbol{P}$, $\boldsymbol{Z}$, $\alpha$, ...) for any one of the metacommunities. (It makes no difference which, since they are identically composed.) Then

$$\boldsymbol{P}^\star = \begin{pmatrix} \frac{1}{m}\boldsymbol{P} & & & \boldsymbol{0} \\ & \frac{1}{m}\boldsymbol{P} & & \\ & & \ddots & \\ \boldsymbol{0} & & & \frac{1}{m}\boldsymbol{P} \end{pmatrix}, \qquad \boldsymbol{Z}^\star = \begin{pmatrix} \boldsymbol{Z} & & & \boldsymbol{0} \\ & \boldsymbol{Z} & & \\ & & \ddots & \\ \boldsymbol{0} & & & \boldsymbol{Z} \end{pmatrix}$$

| | Subcommunity | Metacommunity |
|---|---|---|
| Raw alpha | $\alpha^\star_j = m\alpha_j$ | $A^\star = mA$ |
| Normalised alpha | $\overline{\alpha}^\star_j = \overline{\alpha}_j$ | $\overline{A}^\star = \overline{A}$ |
| Raw beta | $\rho^\star_j = \rho_j$<br>$\beta^\star_j = \beta_j$ | $R^\star = R$<br>$B^\star = B$ |
| Normalised beta | $\overline{\rho}^\star_j = \rho_j/m$<br>$\overline{\beta}^\star_j = m\overline{\beta}_j$ | $\overline{R}^\star = \overline{R}/m$<br>$\overline{B}^\star = m\overline{B}$ |
| Gamma | $\gamma^\star_j = m\gamma_j$ | $G^\star = mG$ |

Table S2.2: Replication (Section S2.1.3): diversity measures for a multicommunity divided into $m$ metacommunities of identical composition.

where both right-hand sides are block sums. Elementary calculations then give the results in Table S2.2.

### S2.1.4 Modularity

More generally, consider a multicommunity divided into metacommunities, each of which is further divided into subcommunities (Figure 1), but no longer assuming that the metacommunities have the same size or composition. We can view the whole multicommunity as divided into subcommunities (ignoring the intermediate metacommunity level), and ask about its alpha-, beta- and gamma-diversities. Can they be expressed in terms of the alpha-, beta- and gamma-diversities within each metacommunity?

The answer is yes, under the (strong) assumption that types in different metacommunities are completely dissimilar.

Suppose that the multicommunity is divided into $m$ metacommunities. For $1 \leq k \leq m$, write $v_k$ for the abundance of the $k$th metacommunity relative to the multicommunity; thus, $\sum_{k=1}^m v_k = 1$. The $k$th metacommunity is further divided into $N_k$ subcommunities. The similarity matrix for each metacommunity need not be naïve, but we assume that the metacommunities have no shared types and that types in different metacommunities are completely dissimilar.

Write $\alpha^k_j$ for the raw subcommunity alpha-diversity of the $j$th subcommunity in the $k$th metacommunity within that metacommunity ($1 \leq j \leq N_k$), write $A^k$ for the raw alpha-diversity of the $k$th metacommunity, etc.

Now view the multicommunity as divided into $N_1 + \cdots + N_m$ subcommunities, ignoring the metacommunity level. The subcommunities are indexed by pairs $(k, j)$ with $1 \leq k \leq m$ and $1 \leq j \leq N_k$. Write $\alpha^{\star k}_j$ for the raw subcommunity alpha-diversity of the $(k, j)$th subcommunity in the multicommunity, write $A^\star$ for the raw multicommunity alpha-diversity, etc.

Elementary calculations give the results shown in Table S2.3, where in the formulae for the multicommunity diversities, we have written $\boldsymbol{A} = (A^1, \ldots, A^k)$, etc.

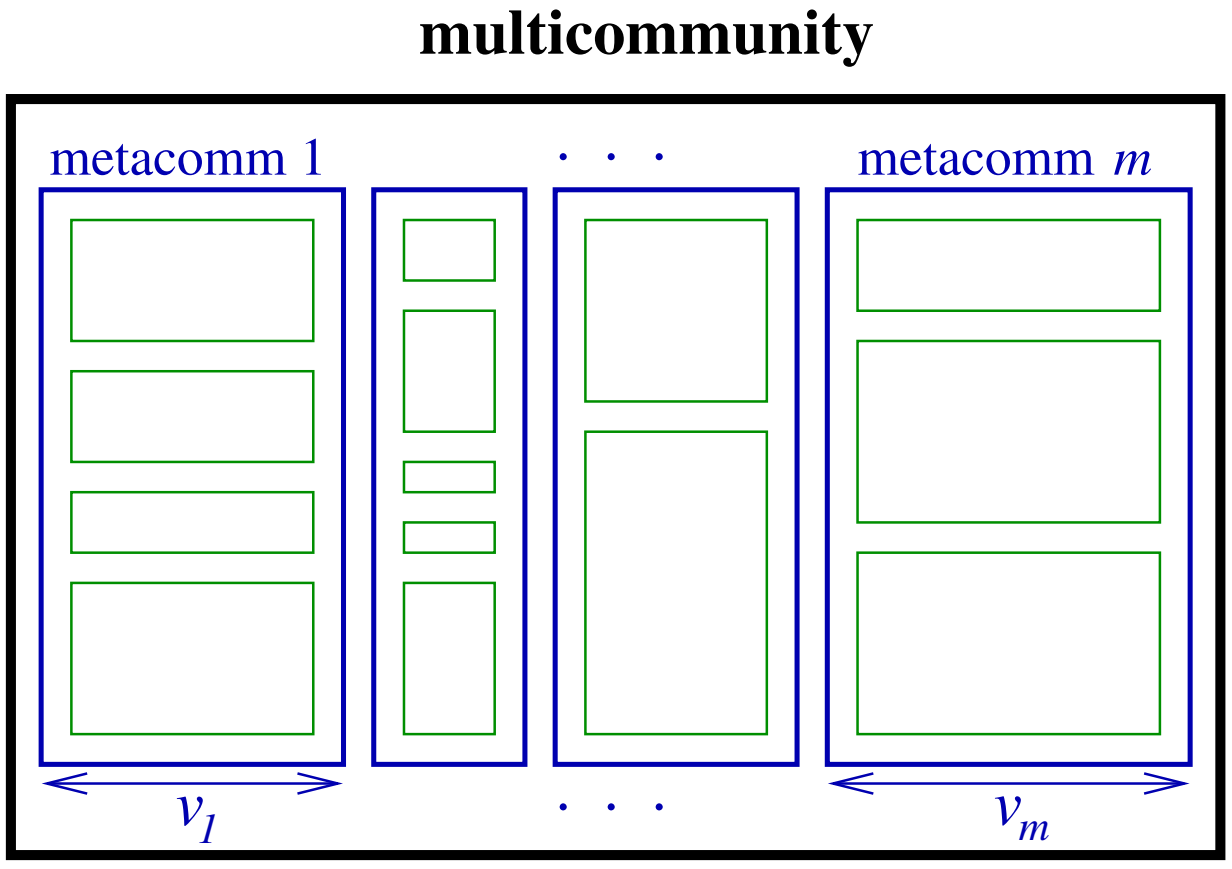

Figure 1: Terminology for the modularity formulae (Section S2.1.4). A multicommunity (black) is divided into metacommunities (blue), each of which is divided further into subcommunities (green). In this example, there are $m = 4$ metacommunities, divided into $N_1 = 4$, $N_2 = 5$, $N_3 = 2$ and $N_4 = 3$ subcommunities, respectively. We may also ignore the metacommunity level and view the multicommunity as divided into $\sum_k N_k = 14$ subcommunities.

| | Subcommunity | Metacommunity |
|---|---|---|
| Raw alpha | $\alpha^{\star k}_{j} = \alpha^k_j / v_k$ | $A^\star = M_{1-q}(\boldsymbol{v}, \boldsymbol{A}/\boldsymbol{v})$ |
| Normalised alpha | $\overline{\alpha}^{\star k}_{j} = \overline{\alpha}^k_j$ | $\overline{A}^\star = M_{1-q}(\boldsymbol{v}, \overline{\boldsymbol{A}})$ |
| Raw beta | $\rho^{\star k}_{j} = \rho^k_j$ <br> $\beta^{\star k}_{j} = \beta^k_j$ | $R^\star = M_{1-q}(\boldsymbol{v}, \boldsymbol{R})$ <br> $B^\star = M_{1-q}(\boldsymbol{v}, \boldsymbol{B})$ |
| Normalised beta | $\overline{\rho}^{\star k}_{j} = v_k \overline{\rho}^k_j$ <br> $\overline{\beta}^{\star k}_{j} = \overline{\beta}^k_j / v_k$ | $\overline{R}^\star = M_{1-q}(\boldsymbol{v}, \boldsymbol{v}\overline{\boldsymbol{R}})$ <br> $\overline{B}^\star = M_{1-q}(\boldsymbol{v}, \overline{\boldsymbol{B}}/\boldsymbol{v})$ |
| Gamma | $\gamma^{\star k}_{j} = \gamma^k_j / v_k$ | $G^\star = M_{1-q}(\boldsymbol{v}, \boldsymbol{G}/\boldsymbol{v})$ |

Table S2.3: Modularity (Section S2.1.4): diversity measures for a multicommunity divided into $m$ metacommunities of relative sizes $v_1, \ldots, v_m$. Here $1 \leq k \leq m$ and $1 \leq j \leq N_k$.

These formulae exhibit two important features of our measures. First, *each diversity measure of a subcommunity within the multicommunity depends only on (i) the corresponding measure for the subcommunity within its metacommunity, and (ii) the size of that metacommunity.* For instance, $\alpha^{\star k}_j$ depends only on $\alpha^k_j$ and $v_k$. Furthermore, the normalised subcommunity alpha-diversity $\overline{\alpha}$, redundancy $\rho$ and distinctiveness $\beta$ are the same whether we view the subcommunity as part of the metacommunity or part of the multicommunity.

Second, *each measure of multicommunity diversity depends only on (i) the corresponding metacommunity diversities and (ii) the sizes of the metacommunities.* For instance, $A^\star$ depends only on $A^1, \ldots, A^m$ and $v_1, \ldots, v_m$; it is independent of the internal structure of the metacommunities.

In the special case where the metacommunities are of equal size and identical structure, we have $\boldsymbol{v} = (1/m, \ldots, 1/m)$ and we recover the replication formulae of Section S2.1.3.

### S2.1.5 Invariance under shattering

Here we prove that the normalised metacommunity diversities $\overline{A}$, $\overline{R}$ and $\overline{B}$ are invariant under shattering (as of course is $G$).

It suffices to prove that $\overline{A}$, $\overline{R}$ and $\overline{B}$ are unchanged when we split a single well-mixed subcommunity into two. Suppose that we split the last subcommunity in proportions $t$ and $1-t$, where $0 < t < 1$. Thus, the new relative abundance matrix $\boldsymbol{P}'$ is the $S \times (N+1)$ matrix with columns

$$\boldsymbol{P}'_{\cdot j} = \begin{cases} \boldsymbol{P}_{\cdot j} & \text{if } j < N \\ t\boldsymbol{P}_{\cdot N} & \text{if } j = N \\ (1-t)\boldsymbol{P}_{\cdot N} & \text{if } j = N+1. \end{cases}$$

Let us write $\boldsymbol{w}'$, $\alpha'_j$, etc., for the analogues of $\boldsymbol{w}$, $\alpha_j$, etc., in the newly-divided metacommunity. Then $\boldsymbol{p}' = \boldsymbol{p}$ and

$$\boldsymbol{w}' = \big(w_1, \ldots, w_{N-1}, tw_N, (1-t)w_N\big), \qquad \overline{\boldsymbol{P}'_{\cdot j}} = \begin{cases} \overline{\boldsymbol{P}_{\cdot j}} & \text{if } j < N \\ \overline{\boldsymbol{P}_{\cdot N}} & \text{if } j \geq N. \end{cases}$$

Now

$$\overline{\alpha}'_j = \begin{cases} \overline{\alpha}_j & \text{if } j < N \\ \overline{\alpha}_N & \text{if } j \geq N \end{cases}$$

(that is, the normalised alpha-diversity of a subcommunity of the new metacommunity is the same as that of its parent in the old metacommunity).

Hence

$$\begin{aligned}\overline{A}' &= M_{1-q}\big((w_1, \ldots, w_{N-1}, tw_N, (1-t)w_N),\ (\overline{\alpha}_1, \ldots, \overline{\alpha}_{N-1}, \overline{\alpha}_N, \overline{\alpha}_N)\big)\\ &= M_{1-q}(\boldsymbol{w}, \overline{\boldsymbol{\alpha}}) = \overline{A},\end{aligned}$$

where the second equality is by an elementary property of power means. Similarly,

$$\overline{\rho}'_j = \begin{cases}\overline{\rho}_j & \text{if } j < N\\ \overline{\rho}_N & \text{if } j \geq N,\end{cases} \qquad \overline{\beta}'_j = \begin{cases}\overline{\beta}_j & \text{if } j < N\\ \overline{\beta}_N & \text{if } j \geq N,\end{cases}$$

giving $\overline{R}' = \overline{R}$ and $\overline{B}' = \overline{B}$ by the same argument as for the alpha-diversities.

The raw metacommunity diversities $A$, $R$ and $B$ are *not* invariant under shattering, as demonstrated by the case of a well-mixed metacommunity (Section S2.2.3); nor would we expect them to be, for the reasons explained in the main text.

### S2.1.6 Independence

Assume that we are using the naïve-type model $\boldsymbol{Z} = \boldsymbol{I}$. In that case, $\overline{A} \geq 1$ and $1 \leq \overline{B} \leq N$ (Section S2.1.2). Here we show that $\overline{A}$ and $\overline{B}$ are independent in the sense of Jost [3], at least for $q \neq 0, \infty$. This means that being told the value of the alpha-diversity $\overline{A}$ gives no certain information about the value of the beta-diversity $\overline{B}$, and vice versa. This is true even if $q$ and $N$ are also assumed to be known.

In precise terms, the assertion is that for all $0 < q < \infty$, integers $N \geq 1$, and real numbers $x$ and $y$ such that $x \geq 1$ and $1 \leq y \leq N$, there exists a metacommunity, divided into $N$ subcommunities, such that $\overline{A} = x$ and $\overline{B} = y$.

One way of constructing such a system is as follows. Choose an integer $S \geq x$. For probability vectors $\boldsymbol{t} = (t_1, \ldots, t_S)$, the diversity ${}^qD(\boldsymbol{t})$ is continuous in $\boldsymbol{t}$, and has maximum value $S$ and minimum value 1. (The statement on continuity requires $q \neq 0, \infty$; see Proposition A2 of the appendix of [4].) Hence we may choose $\boldsymbol{t}$ such that ${}^qD(\boldsymbol{t}) = x$. Similarly, we may choose $\boldsymbol{w} = (w_1, \ldots, w_N)$ such that ${}^qD(\boldsymbol{w}) = y$.

Now consider a metacommunity made up of $N$ subcommunities of relative sizes $w_1, \ldots, w_N$, with no shared types, where each subcommunity contains $S$ types in proportions $t_1, \ldots, t_S$. (Thus, there are $NS$ types in all.) The diversity of each subcommunity is ${}^qD(\boldsymbol{t}) = x$; that is, $\overline{\alpha}_j = x$ for all $j$. Hence $\overline{A} = x$. Moreover, the results of Section S2.2.4 imply that $\overline{B} = {}^qD(\boldsymbol{w}) = y$, as required.

In the same sense, $\overline{A}$ and $R$ are independent. To prove this, fix $0 < q < \infty$ and an integer $N \geq 1$. We have already noted that $\overline{A} \geq 1$, and also $1 \leq R \leq N$ (Section S2.1.2, continuing to assume that $\boldsymbol{Z} = \boldsymbol{I}$). Now take any real numbers $x$ and $z$ with $x \geq 1$ and $1 \leq z \leq N$. Choose an integer $S \geq x$, then choose probability vectors $\boldsymbol{p} = (p_1, \dots, p_S)$ and $\boldsymbol{w} = (w_1, \dots, w_N)$ such that ${}^qD(\boldsymbol{p}) = x$ and ${}^qD(\boldsymbol{w}) = z$. Consider a metacommunity made up of $N$ subcommunities of relative sizes $w_1, \dots, w_N$, each containing the same $S$ types in proportions $p_1, \dots, p_S$. Then by the results of Section S2.2.3, $\overline{A} = {}^qD(\boldsymbol{p}) = x$ and $R = {}^qD(\boldsymbol{w}) = z$, as required.

### S2.1.7 Jost's measures

Operating within the naïve-type model $\boldsymbol{Z} = \boldsymbol{I}$, Jost [2] defines metacommunity alpha-, beta- and gamma-diversities, all depending on a parameter $q$ ($0 \leq q < \infty$). His gamma-diversity is the same as our $G$; it is simply the metacommunity Hill number. His alpha- and beta-diversities are most closely analogous to our $\overline{A}$ and $\overline{B}$, so we denote them here by $\overline{A}_{\text{Jost}}$ and $\overline{B}_{\text{Jost}}$. They are defined by

$$\overline{A}_{\text{Jost}} = M_{1-q}(\boldsymbol{w}^{(q)}, \overline{\boldsymbol{\alpha}}), \qquad \overline{B}_{\text{Jost}} = G/\overline{A}_{\text{Jost}},$$

where $\boldsymbol{w}^{(q)} = (w_1^{(q)}, \dots, w_N^{(q)})$ and $w_j^{(q)} = w_j^q / \sum_{j'} w_{j'}^q$.

**Comparison with our measures** Since $\overline{A} = M_{1-q}(\boldsymbol{w}, \overline{\boldsymbol{\alpha}})$ and $\boldsymbol{w}^{(q)} \neq \boldsymbol{w}$ in general, the measures $\overline{A}_{\text{Jost}}$ and $\overline{A}$ are usually not equal. But we do have $\boldsymbol{w}^{(1)} = \boldsymbol{w}$, so $\overline{A}_{\text{Jost}} = \overline{A}$ when $q = 1$. Moreover, it is shown in Section S2.2.5 that when $q = 1$ in the naïve-type model, $\overline{B} = G/\overline{A}$; so in that case, $\overline{B}_{\text{Jost}} = \overline{B}$ too.

Hence when $q = 1$, in the naïve-type model, our $\overline{A}$, $\overline{B}$ and $G$ reproduce Jost's alpha-, beta- and gamma-diversities.

For arbitrary $q < \infty$, if the subcommunities are of equal size ($\boldsymbol{w} = (1/N, \dots, 1/N)$) or of equal diversity ($\overline{\alpha}_1 = \dots = \overline{\alpha}_N$) then $\overline{A}_{\text{Jost}} = \overline{A}$, but even so, $\overline{B}_{\text{Jost}} \neq \overline{B}$.

**Failure of invariance under shattering** Jost's measures are not invariant under shattering unless $q = 1$ (in which case they coincide with ours). For example, take a metacommunity divided into two equal-sized subcommunities with respective diversities $d_1$ and $d_2$, where $d_1 \neq d_2$. Then for $q \neq 1$,

$$\begin{aligned}\overline{A}_{\text{Jost}} &= M_{1-q}\big((1/2, 1/2)^{(q)}, (d_1, d_2)\big) \\ &= \Big((1/2)d_1^{1-q} + (1/2)d_2^{1-q}\Big)^{1/(1-q)}.\end{aligned}$$

Now split the second subcommunity (which we suppose to be well-mixed) into two equal-sized parts. In this newly-divided metacommunity,

$$\begin{aligned}\overline{A}_{\text{Jost}} &= M_{1-q}\big((1/2, 1/4, 1/4)^{(q)}, (d_1, d_2, d_2)\big) \\ &= \left(\frac{(1/2)^q}{(1/2)^q + 2(1/4)^q} d_1^{1-q} + \frac{2(1/4)^q}{(1/2)^q + 2(1/4)^q} d_2^{1-q}\right)^{1/(1-q)}.\end{aligned}$$

The two values of $\overline{A}_{\text{Jost}}$, before and after shattering, are not equal. Since $G$ is unchanged, the two values of $\overline{B}_{\text{Jost}} = G/\overline{A}_{\text{Jost}}$ are not equal either. Hence for $q \neq 1$, neither $\overline{A}_{\text{Jost}}$ nor $\overline{B}_{\text{Jost}}$ is invariant under shattering.

## S2.2 Examples

Most of the examples that follow are hypothetical extreme cases, intended to shed light on the measures that we have defined.

### S2.2.1 Metacommunity is a single subcommunity

First suppose that our metacommunity is divided into a single subcommunity: that is, not divided at all. Then $N = 1$, $\boldsymbol{P}_{\cdot 1} = \overline{\boldsymbol{P}_{\cdot 1}} = \boldsymbol{p}$, $\boldsymbol{w} = (1)$, and

$$\begin{aligned}\alpha_1 = \overline{\alpha}_1 = A = \overline{A} &= {}^qD^{\boldsymbol{Z}}(\boldsymbol{p}), \\ \rho_1 = \beta_1 = \overline{\rho}_1 = \overline{\beta}_1 = R = B = \overline{R} = \overline{B} &= 1, \\ \gamma_1 = G &= {}^qD^{\boldsymbol{Z}}(\boldsymbol{p}).\end{aligned}$$

### S2.2.2 Subcommunities are types

At the opposite extreme, suppose that the subcommunities are exactly the types. Then $S = N$, $\boldsymbol{P}$ is the diagonal matrix with entries $p_1, \ldots, p_S$, $\boldsymbol{w} = \boldsymbol{p}$, $\boldsymbol{Z}\overline{\boldsymbol{P}_{\cdot j}} = \boldsymbol{Z}_{\cdot j}$, and in particular, $(\boldsymbol{Z}\overline{\boldsymbol{P}_{\cdot j}})_j = 1$ (since every type is assumed to have a self-similarity of 1). Straightforward calculations give the results in Table S2.4. Note that $A$ is the naïve diversity ${}^qD(\boldsymbol{p})$, not ${}^qD^{\boldsymbol{Z}}(\boldsymbol{p})$.

### S2.2.3 Well-mixed metacommunity

Take an arbitrary metacommunity, with relative abundance vector $\boldsymbol{p} = (p_1, \ldots, p_S)$ and similarity matrix $\boldsymbol{Z}$. Assume that the metacommunity is well-mixed, so that no matter how we divide it geographically into subcommunities, all those subcommunities will have the same distribution of types.

In mathematical terms, we take any $N \geq 1$ and any probability vector $\boldsymbol{w} = (w_1, \ldots, w_N)$. Put $P_{ij} = p_i w_j$. Then $\boldsymbol{P} = (P_{ij})$ is the relative abundance matrix for our metacommunity divided into subcommunities of sizes $w_1, \ldots, w_N$. We have $\overline{\boldsymbol{P}_{\cdot j}} = \boldsymbol{p}$ for all $j$, and the measures of our divided metacommunity are as shown in Table S2.5.

| | Subcommunity | Metacommunity |
|---|---|---|
| Raw alpha | $\alpha_j = 1/p_j$ | $A = {}^qD(\boldsymbol{p})$ |
| Normalised alpha | $\overline{\alpha}_j = 1$ | $\overline{A} = 1$ |
| Raw beta | $\rho_j = (\boldsymbol{Zp})_j/p_j$<br>$\beta_j = p_j/(\boldsymbol{Zp})_j$ | $R = M_{1-q}(\boldsymbol{p}, \boldsymbol{Zp}/\boldsymbol{p})$<br>$B = M_{1-q}(\boldsymbol{p}, \boldsymbol{p}/\boldsymbol{Zp})$ |
| Normalised beta | $\overline{\rho}_j = (\boldsymbol{Zp})_j$<br>$\overline{\beta}_j = 1/(\boldsymbol{Zp})_j$ | $\overline{R} = M_{1-q}(\boldsymbol{p}, \boldsymbol{Zp})$<br>$\overline{B} = {}^qD^{\boldsymbol{Z}}(\boldsymbol{p})$ |
| Gamma | $\gamma_j = 1/(\boldsymbol{Zp})_j$ | $G = {}^qD^{\boldsymbol{Z}}(\boldsymbol{p})$ |

Table S2.4: Diversity measures when the subcommunities are exactly the types (Section S2.2.2).

| | Subcommunity | Metacommunity |
|---|---|---|
| Raw alpha | $\alpha_j = {}^qD^{\boldsymbol{Z}}(\boldsymbol{p})/w_j$ | $A = {}^qD^{\boldsymbol{Z}}(\boldsymbol{p}) \cdot {}^qD(\boldsymbol{w})$ |
| Normalised alpha | $\overline{\alpha}_j = {}^qD^{\boldsymbol{Z}}(\boldsymbol{p})$ | $\overline{A} = {}^qD^{\boldsymbol{Z}}(\boldsymbol{p})$ |
| Raw beta | $\rho_j = 1/w_j$<br>$\beta_j = w_j$ | $R = {}^qD(\boldsymbol{w})$<br>$B = M_{1-q}(\boldsymbol{w}, \boldsymbol{w})$ |
| Normalised beta | $\overline{\rho}_j = 1$<br>$\overline{\beta}_j = 1$ | $\overline{R} = 1$<br>$\overline{B} = 1$ |
| Gamma | $\gamma_j = {}^qD^{\boldsymbol{Z}}(\boldsymbol{p})$ | $G = {}^qD^{\boldsymbol{Z}}(\boldsymbol{p})$ |

Table S2.5: Diversity measures for a well-mixed metacommunity (Section S2.2.3).

Note that the normalised measures $\overline{A}$, $\overline{R}$ and $\overline{B}$ are independent of $\boldsymbol{w}$ (as of course is $G$). This is a special case of invariance under shattering. But the raw measures $A$, $R$ and $B$ clearly do depend on $\boldsymbol{w}$, so they are not invariant under shattering.

### S2.2.4 Naïve-community model

Take a metacommunity divided into $N$ subcommunities with nothing at all in common, in the sense that (i) no type is present in more than one subcommunity, and (ii) the types present in each subcommunity are completely dissimilar from the types present in the others.

Label the types in the first subcommunity as $1, \ldots, S_1$, those in the second as $S_1 + 1, \ldots, S_1 + S_2$, and so on. Then the total number of types is

| | Subcommunity | Metacommunity |
|---|---|---|
| Raw alpha | $\alpha_j = \overline{\alpha}_j / w_j$ | $A = M_{1-q}(\boldsymbol{w}, \overline{\boldsymbol{\alpha}}/\boldsymbol{w})$ |
| Normalised alpha | $\overline{\alpha}_j = {}^qD^{\boldsymbol{Z}^j}(\boldsymbol{t}_{\cdot j}/w_j)$ | $\overline{A} = M_{1-q}(\boldsymbol{w}, \overline{\boldsymbol{\alpha}})$ |
| Raw beta | $\rho_j = 1$<br>$\beta_j = 1$ | $R = 1$<br>$B = 1$ |
| Normalised beta | $\overline{\rho}_j = w_j$<br>$\overline{\beta}_j = 1/w_j$ | $\overline{R} = M_{1-q}(\boldsymbol{w}, \boldsymbol{w})$<br>$\overline{B} = {}^qD(\boldsymbol{w})$ |
| Gamma | $\gamma_j = \overline{\alpha}_j / w_j$ | $G = M_{1-q}(\boldsymbol{w}, \overline{\boldsymbol{\alpha}}/\boldsymbol{w})$ |

Table S2.6: Diversity measures in the naïve-community case (Section S2.2.4).

$S = S_1 + \cdots + S_N$, and the distribution matrix $\boldsymbol{P}$ has columns of the form

$$\boldsymbol{P}_{\cdot j} = (\underbrace{0, \ldots, 0}_{S_1 + \cdots + S_{j-1}}, t_{1j}, \ldots, t_{S_j j}, \underbrace{0, \ldots, 0}_{S_{j+1} + \cdots + S_N})$$

(or strictly speaking, the transpose of this vector). Hence

$$\begin{aligned} \boldsymbol{p} &= (t_{11}, \ldots, t_{S_1 1}, \quad \ldots, \quad t_{1N}, \ldots, t_{S_N N}), \\ \boldsymbol{w} &= (t_{11} + \cdots + t_{S_1 1}, \quad \ldots, \quad t_{1N} + \cdots + t_{S_N N}). \end{aligned}$$

The similarity matrix $\boldsymbol{Z}$ decomposes as a block sum

$$\boldsymbol{Z} = \begin{pmatrix} \boldsymbol{Z}^1 & & & \boldsymbol{0} \\ & \boldsymbol{Z}^2 & & \\ & & \ddots & \\ \boldsymbol{0} & & & \boldsymbol{Z}^N \end{pmatrix}$$

where $\boldsymbol{Z}^j$ is an $S_j \times S_j$ matrix.

In Table S2.6, the measures of the divided metacommunity are shown in terms of the diversities $\overline{\alpha}_1, \ldots, \overline{\alpha}_N$ of the various subcommunities considered in isolation. In the expression for $\overline{\alpha}_j$, the vector $\boldsymbol{t}_{\cdot j}$ is $(t_{1j}, \ldots, t_{S_j j})$ (and $\boldsymbol{t}_{\cdot j}/w_j$ is its normalisation).

The metacommunity measures have several notable features. First, the beta-diversities $B = 1$ and $\overline{B} = {}^qD(\boldsymbol{w})$ attain their maximum possible values subject to the given $\boldsymbol{w}$ (as derived in Section S2.1.2). This is because the subcommunities have nothing in common. Second, and for the same reason, $G = A$; that is, equality is achieved in the general inequality $G \leq A$. Third, $G = AB$, which is atypical. (Compare Section S2.1.7.) Finally, the formula for $G$ is precisely the modularity formula proved in Proposition A10 of Leinster and Cobbold [4].

### S2.2.5 Naïve-type model

Now suppose that we are using the naïve-type model $\boldsymbol{Z} = \boldsymbol{I}$. In that case, the formulae for most of the metacommunity measures simplify considerably.

In the equations that follow, we view the whole matrix $\boldsymbol{P} = (P_{ij})_{1\leq i\leq S,\, 1\leq j\leq N}$ as an $SN$-dimensional probability vector, and treat the vectors $\boldsymbol{p}$ and $\boldsymbol{w}$ as $SN$-dimensional too, with $(i,j)$-entries $p_i$ and $w_j$ respectively. We also define an $SN$-dimensional vector $\boldsymbol{p}\otimes\boldsymbol{w}$ with $(i,j)$-entry $p_i w_j$. In this notation,

$$
\begin{aligned}
A &= M_{1-q}(\boldsymbol{P}, 1/\boldsymbol{P}) = {}^{q}D(\boldsymbol{P}), & \overline{A} &= M_{1-q}(\boldsymbol{P}, \boldsymbol{w}/\boldsymbol{P}),\\
R &= M_{1-q}(\boldsymbol{P}, \boldsymbol{p}/\boldsymbol{P}), & \overline{R} &= M_{1-q}(\boldsymbol{P}, (\boldsymbol{p}\otimes\boldsymbol{w})/\boldsymbol{P}),\\
G &= M_{1-q}(\boldsymbol{p}, 1/\boldsymbol{p}) = {}^{q}D(\boldsymbol{p}).
\end{aligned}
$$

We recover two of the quantities described in Tuomisto's survey of beta-diversities [5]. First, $\overline{A}$ is the quantity that Tuomisto denotes by $\alpha_{\mathrm{t}}$, the 't' standing for 'turnover'. Second, when there are no shared types between subcommunities, the calculations in Section S2.2.4 give $\overline{B} = {}^{q}D(\boldsymbol{w})$, which is what Tuomisto denotes by ${}^{q}D_{\omega}$.

Table S2.7 shows the behaviour of our measures in the naïve-type model for $q = 0, 1, \infty$, respectively.

For $q = 0$, we adopt the following notation: given a vector $\boldsymbol{v} = (v_1, \ldots, v_n)$, we write $\#\boldsymbol{v}$ for the number of indices $i$ such that $v_i \neq 0$. The table shows, for instance, that $\overline{A}$ is simply the average number of types present in a subcommunity, weighted according to subcommunity size. The normalised subcommunity beta-diversity $\overline{\beta}_j$ is highest when the $j$th subcommunity consists exclusively of types that are rare in the metacommunity, but decreases sharply if it contains even one individual of a common type.

For $q = 1$, we have the multiplicative relationships $AB = G = \overline{A}\,\overline{B}$, and our measures $\overline{A}$ and $\overline{B}$ coincide with Jost's alpha- and beta-diversities. See also Sections S2.1.7 and S2.2.6.

For $q = \infty$, the distinctiveness $\beta_j$ of the $j$th subcommunity is high if there is at least one type that is confined almost entirely to that subcommunity. This conforms with the intuitive idea that $\beta_j$ is high if the $j$th subcommunity is in some sense unusual relative to the metacommunity. The metacommunity measure $B$ is high if every subcommunity is home to at least one such specialist type.

### S2.2.6 Arbitrary similarity matrix with $q = 1$

In the previous section, we computed the diversity measures of a divided community when $\boldsymbol{Z} = \boldsymbol{I}$ and $q = 1$, noting the multiplicative relationships $AB = G = \overline{A}\,\overline{B}$. In fact, these relationships persist when $q = 1$ for an *arbitrary* similarity matrix $\boldsymbol{Z}$, as shown in Table S2.8. They also hold at the subcommunity level: $\alpha_j\beta_j = \gamma_j = \overline{\alpha}_j\overline{\beta}_j$.

| $q=0$ | Subcommunity | Metacommunity |
|---|---|---|
| Raw alpha | $\alpha_j = \frac{\#\boldsymbol{P}_{\cdot j}}{w_j}$ | $A = \#\boldsymbol{P}$ |
| Normalised alpha | $\overline{\alpha}_j = \#\boldsymbol{P}_{\cdot j}$ | $\overline{A} = \sum_j w_j \cdot \#\boldsymbol{P}_{\cdot j}$ |
| Raw beta | $\rho_j = \frac{1}{w_j} \sum_{i\colon P_{ij}>0} p_i$ <br> $\beta_j = 1/\rho_j$ | $R = \sum_i p_i \cdot \#\boldsymbol{P}_{i\cdot}$ <br> $B = \sum_j \frac{w_j^2}{\sum_{i\colon P_{ij}>0} p_i}$ |
| Normalised beta | $\overline{\rho}_j = \sum_{i\colon P_{ij}>0} p_i$ <br> $\overline{\beta}_j = 1/\overline{\rho}_j$ | $\overline{R} = \sum_{i,j\colon P_{ij}>0} p_i w_j$ <br> $\overline{B} = \sum_j \frac{w_j}{\sum_{i\colon P_{ij}>0} p_i}$ |
| Gamma | $\gamma_j = \sum_{i,j\colon P_{ij}>0} \frac{P_{ij}}{p_i w_j}$ | $G = \#\boldsymbol{p}$ |

| $q=1$ | Subcommunity | Metacommunity |
|---|---|---|
| Raw alpha | $\alpha_j = \prod_i P_{ij}^{-\overline{P_{ij}}}$ | $A = \prod_{i,j} P_{ij}^{-P_{ij}}$ |
| Normalised alpha | $\overline{\alpha}_j = \prod_i \overline{P_{ij}}^{-\overline{P_{ij}}}$ | $\overline{A} = \prod_{i,j} \overline{P_{ij}}^{-P_{ij}}$ |
| Raw beta | $\rho_j = \alpha_j/\gamma_j$ <br> $\beta_j = \gamma_j/\alpha_j$ | $R = A/G$ <br> $B = G/A$ |
| Normalised beta | $\overline{\rho}_j = \overline{\alpha}_j/\gamma_j$ <br> $\overline{\beta}_j = \gamma_j/\overline{\alpha}_j$ | $\overline{R} = \overline{A}/G$ <br> $\overline{B} = G/\overline{A}$ |
| Gamma | $\gamma_j = \prod_i p_i^{-\overline{P_{ij}}}$ | $G = \prod_i p_i^{-p_i}$ |

| $q=\infty$ | Subcommunity | Metacommunity |
|---|---|---|
| Raw alpha | $\alpha_j = 1/\max_{i\colon P_{ij}>0} P_{ij}$ | $A = 1/\max_{i,j\colon P_{ij}>0} P_{ij}$ |
| Normalised alpha | $\overline{\alpha}_j = 1/\max_{i\colon P_{ij}>0} \overline{P_{ij}}$ | $\overline{A} = 1/\max_{i,j\colon P_{ij}>0} \overline{P_{ij}}$ |
| Raw beta | $\rho_j = \min_{i\colon P_{ij}>0} p_i/P_{ij}$ <br> $\beta_j = \max_{i\colon P_{ij}>0} P_{ij}/p_i$ | $R = \min_{i,j\colon P_{ij}>0} p_i/P_{ij}$ <br> $B = \min_j \max_{i\colon P_{ij}>0} P_{ij}/p_i$ |
| Normalised beta | $\overline{\rho}_j = \min_{i\colon P_{ij}>0} p_i/\overline{P_{ij}}$ <br> $\overline{\beta}_j = \max_{i\colon P_{ij}>0} \overline{P_{ij}}/p_i$ | $\overline{R} = \min_{i,j\colon P_{ij}>0} p_i/\overline{P_{ij}}$ <br> $\overline{B} = \min_j \max_{i\colon P_{ij}>0} \overline{P_{ij}}/p_i$ |
| Gamma | $\gamma_j = 1/\max_{i\colon P_{ij}>0} p_i$ | $G = 1/\max_{i\colon p_i>0} p_i$ |

Table S2.7: Diversity measures in the naïve-type model $\boldsymbol{Z} = \boldsymbol{I}$ for $q = 0$, 1 and $\infty$ (Section S2.2.5). For the definition of #, see the text.

| | Subcommunity | Metacommunity |
|---|---|---|
| Raw alpha | $\alpha_j = \prod_i (\boldsymbol{ZP}_{\cdot j})_i^{-\overline{P_{ij}}}$ | $A = \prod_{i,j} (\boldsymbol{ZP}_{\cdot j})_i^{-P_{ij}}$ |
| Normalised alpha | $\overline{\alpha}_j = \prod_i (\boldsymbol{Z}\overline{\boldsymbol{P}_{\cdot j}})_i^{-\overline{P_{ij}}}$ | $\overline{A} = \prod_{i,j} (\boldsymbol{Z}\overline{\boldsymbol{P}_{\cdot j}})_i^{-P_{ij}}$ |
| Raw beta | $\rho_j = \alpha_j / \gamma_j$ <br> $\beta_j = \gamma_j / \alpha_j$ | $R = A/G$ <br> $B = G/A$ |
| Normalised beta | $\overline{\rho}_j = \overline{\alpha}_j / \gamma_j$ <br> $\overline{\beta}_j = \gamma_j / \overline{\alpha}_j$ | $\overline{R} = \overline{A}/G$ <br> $\overline{B} = G/\overline{A}$ |
| Gamma | $\gamma_j = \prod_i (\boldsymbol{Zp})_i^{-\overline{P_{ij}}}$ | $G = \prod_i (\boldsymbol{Zp})_i^{-p_i}$ |

Table S2.8: Diversity measures for arbitrary $\boldsymbol{Z}$ and $q = 1$ (Section S2.2.6).

### S2.2.7 Distinctiveness and turnover

Distinctiveness ($\beta_j$) can be understood as a kind of turnover between subommunity $j$ and the rest of the metacommunity (see main text, *Distinctiveness*). For example, consider a naïve-type case ($\boldsymbol{Z} = \boldsymbol{I}$) with $N$ subcommunities ordered along a spatial gradient, with *absolute* abundance matrix

$$\begin{pmatrix}
1 & & & & & \\
1 & & & & & \\
1 & 1 & & & & \\
1 & 1 & & & & \\
1 & 1 & 1 & & & \\
1 & 1 & 1 & & & \\
 & 1 & 1 & 1 & & \\
 & 1 & 1 & 1 & & \\
 & & 1 & 1 & & \\
 & & 1 & 1 & \ddots & \\
 & & & 1 & & 1 \\
 & & & 1 & & 1 \\
 & & & & \ddots & 1 \\
 & & & & & 1 \\
 & & & & & 1 \\
 & & & & & 1
\end{pmatrix}$$

where unmarked entries are zero. Thus, each subcommunity contains 6 individuals, one of each of 6 types; there is a change of 1/3 of the types from each subcommunity to the next. There are $6N$ individuals in total, so for

$3 \leq j \leq N-2$, we have

$$\boldsymbol{p} = \begin{pmatrix} 1/6N \\ 1/6N \\ 2/6N \\ 2/6N \\ 3/6N \\ 3/6N \\ \vdots \\ 3/6N \\ 2/6N \\ 2/6N \\ 1/6N \\ 1/6N \end{pmatrix}, \qquad \boldsymbol{P}_{\cdot j} = \begin{pmatrix} 0 \\ \vdots \\ 0 \\ 1/6N \\ 1/6N \\ 1/6N \\ 1/6N \\ 1/6N \\ 1/6N \\ 0 \\ \vdots \\ 0 \end{pmatrix}, \qquad \boldsymbol{p}/\boldsymbol{P}_{\cdot j} = \begin{pmatrix} \infty \\ \vdots \\ \infty \\ 3 \\ 3 \\ 3 \\ 3 \\ 3 \\ 3 \\ \infty \\ \vdots \\ \infty \end{pmatrix}.$$

(We restrict to $3 \leq j \leq N-2$ to avoid end effects. For instance, although most types are present in exactly three subcommunities, the first and last two types are only present in one.) So for $3 \leq j \leq N-2$,

$$\begin{aligned} \rho_j &= M_{1-q}(\overline{\boldsymbol{P}_{\cdot j}}, \boldsymbol{p}/\boldsymbol{P}_{\cdot j}) \\ &= M_{1-q}\big((1/6, 1/6, 1/6, 1/6, 1/6, 1/6),\ (3, 3, 3, 3, 3, 3)\big) \\ &= 3. \end{aligned}$$

It follows that $\beta_j = 1/\rho_j = 1/3$ for $3 \leq j \leq N-2$, as claimed in the text. For $j < 3$ or $j > N-2$, the redundancy $\rho_j$ is a mean of values in the set

$$\left\{ \frac{1/6N}{1/6N}, \frac{2/6N}{1/6N}, \frac{3/6N}{1/6N} \right\} = \{1, 2, 3\},$$

so $1 \leq \rho_j \leq 3$ and $1/3 \leq \beta_j \leq 1$. Note that for $j < 3$, both $\beta_j$ and $\beta_{N+1-j}$ are independent of $N$.

To calculate $B$: we have

$$\begin{aligned} B &= M_{1-q}\big((\tfrac{1}{N}, \ldots, \tfrac{1}{N}),\ (\beta_1, \beta_2, \tfrac{1}{3}, \ldots, \tfrac{1}{3}, \beta_{N-1}, \beta_N)\big) \\ &= M_{1-q}\big((\tfrac{1}{N}, \tfrac{1}{N}, \tfrac{N-4}{N}, \tfrac{1}{N}, \tfrac{1}{N}),\ (\beta_1, \beta_2, \tfrac{1}{3}, \beta_{N-1}, \beta_N)\big) \\ &\to M_{1-q}\big((0, 0, 1, 0, 0),\ (\beta_1, \beta_2, \tfrac{1}{3}, \beta_{N-1}, \beta_N)\big) \\ &= \tfrac{1}{3} \end{aligned}$$

as $N \to \infty$. Here the second equality follows from elementary properties of power means, and the convergence follows from (i) the fact that for $j < 3$, both $\beta_j$ and $\beta_{N+1-j}$ are independent of $N$, and (ii) the continuity of power means in their first argument. This continuity only holds for power means of order not equal to $\pm\infty$, so the case $q = \infty$ needs separate consideration. In that case,

$$B = \min\{\beta_1, \beta_2, \tfrac{1}{3}, \beta_{N-1}, \beta_N\} = \tfrac{1}{3}$$

for all $N$, by the earlier observation that $\beta_j \geq 1/3$ whenever $j < 3$ or $j > N-2$. In summary: for all $0 \leq q \leq \infty$, we have $B \to 1/3$ as $N \to \infty$, as claimed.

### S2.2.8 Representativeness

Consider a naïve-type situation where all types are equally abundant in the metacommunity and equally abundant in each subcommunity in which they are present. Suppose that every subcommunity contains a proportion $r$ of the total number of types. It is claimed in the main text (*Representativeness*) that $\overline{\rho}_j = \overline{R} = r$ for all $j$. Here we prove this.

We are assuming, first, that $\boldsymbol{Z} = \boldsymbol{I}$ and $\boldsymbol{p} = (1/S, \ldots, 1/S)$. Writing $r = s/S$ (where $s \in \{1, \ldots, S\}$), we are also assuming that each subcommunity contains exactly $s$ types, and that they are equally abundant in the subcommunity. Let $1 \leq j \leq N$. By reordering the types if necessary, we may suppose that

$$\overline{\boldsymbol{P}_{\cdot j}} = (\underbrace{1/s, \ldots, 1/s}_{s}, 0, \ldots, 0).$$

So the representativeness of subcommunity $j$ is

$$\overline{\rho}_j = M_{1-q}(\overline{\boldsymbol{P}_{\cdot j}}, \boldsymbol{p}/\overline{\boldsymbol{P}_{\cdot j}}) = M_{1-q}\big((1/s, \ldots, 1/s), (s/S, \ldots, s/S)\big) = s/S = r,$$

as claimed. This holds for all $j$, so $\overline{R} = r$ too.

### S2.2.9 Effect of subcommunity size on gamma diversity

Here we analyse the first of the two examples from the *Gamma diversity* section of the main text:

> Consider, for example, a naïve-type case with all types equally abundant in the metacommunity. If two communities have evenly distributed types, but the first contains $k$ times as many types as the second, then $\overline{\alpha}_1 = k\overline{\alpha}_2$; but $\gamma_1 = \gamma_2$.

We are supposing, then, that $\boldsymbol{Z} = \boldsymbol{I}$ and $\boldsymbol{p} = (1/S, \ldots, 1/S)$. If the first subcommunity contains $s$ types then the second contains $ks$ types, and after reordering the types if necessary, we therefore have

$$\overline{\boldsymbol{P}_{\cdot 1}} = (0, \ldots, 0, \underbrace{1/ks, \ldots, 1/ks}_{ks}, 0, \ldots, 0),$$
$$\overline{\boldsymbol{P}_{\cdot 2}} = (0, \ldots, 0, \underbrace{1/s, \ldots, 1/s}_{s}, 0, \ldots, 0).$$

(It makes no difference how many zeroes there are at the beginning and end of these vectors.) Then

$$\overline{\alpha}_1 = {}^qD(1/ks, \ldots, 1/ks) = ks, \qquad \overline{\alpha}_2 = {}^qD(1/s, \ldots, 1/s) = s,$$

so $\overline{\alpha}_1 = k\overline{\alpha}_2$. Also,

$$\gamma_1 = M_{1-q}(\overline{\boldsymbol{P}_{\cdot 1}}, 1/\boldsymbol{p}) = M_{1-q}\big((1/ks, \ldots, 1/ks), (S, \ldots, S)\big) = S$$

and similarly $\gamma_2 = S$, so $\gamma_1 = \gamma_2$, as claimed.

### S2.2.10 Subcommunities consisting of only rare types

Our subcommunity measures are illuminated by considering a subcommunity consisting entirely of types that are rare within the metacommunity. In this case, $\gamma_j$ is high, even though $\overline{\alpha}_j$, the diversity of the subcommunity in isolation, may be low. For example, when the subcommunity consists of a single rare type, both $\gamma_j$ and $\overline{\beta}_j$ are high, while $\overline{\alpha}_j = 1$. However, if the subcommunity is expanded to include several more types of the same rarity, then $\gamma_j$ continues to be high, since the amount that each individual in the subcommunity contributes to metacommunity diversity is unchanged; $\overline{\beta}_j$ decreases, since the presence of more types makes the type distribution of the subcommunity closer to that of the metacommunity; and $\overline{\alpha}_j$ increases, since there are more types.

For a specific example, let us work in both the naïve-type model $\boldsymbol{Z} = \boldsymbol{I}$ (different types are completely dissimilar) and the naïve-community model (different subcommunities share no types). Suppose further that each type present in either of the first two subcommunities has a relative abundance of $10^{-9}$ in the metacommunity. The first subcommunity consists entirely of one such type, and the second consists of ten such types in equal proportions.

The subcommunity gamma-diversities are $\gamma_1 = \gamma_2 = 10^9$ (for all $q$), by a calculation similar to that in Section S2.2.9. This high value reflects the fact that each individual in either of these subcommunities contributes a great deal to the metacommunity diversity. It is unaffected by how many types the subcommunity contains.

On the other hand, $\overline{\beta}_1 = 10^9 > \overline{\beta}_2 = 10^8$. Both values are high, because both subcommunities have relative abundance distributions highly unlike that of the whole metacommunity; but the first is higher, because the contrast is even greater for the first subcommunity than the second.

Finally, $\overline{\alpha}_1 = 1 < \overline{\alpha}_2 = 10$, reflecting the greater diversity of the second subcommunity when taken in isolation.

### S2.2.11 Effect of rarity on gamma diversity

Here we analyse the second of the two examples from the *Gamma diversity* section of the main text:

> if two equally sized subcommunities have different constituent types, but with the same relative abundances, in such a way that all of the types in the first subcommunity are $k$ times rarer in the metacommunity than the types in the second, then $\overline{\alpha}_1 = \overline{\alpha}_2$ but $\gamma_1 = k\gamma_2$.

Here we are using the naïve-type model $\boldsymbol{Z} = \boldsymbol{I}$.

Suppose that the first two subcommunities have $s$ types each. Then for some nonnegative real numbers $v_1, \ldots, v_s$ summing to 1, we have

$$\overline{\boldsymbol{P}_{\cdot 1}} = (v_1, \ldots, v_s, \underbrace{0, \ldots, 0}_{s}, \underbrace{0, \ldots, 0}_{S-2s}),$$
$$\overline{\boldsymbol{P}_{\cdot 2}} = (\underbrace{0, \ldots, 0}_{s}, v_1, \ldots, v_s, \underbrace{0, \ldots, 0}_{S-2s}),$$
$$\boldsymbol{p} = (p_1, \ldots, p_s, kp_1, \ldots, kp_s, p_{2s+1}, \ldots, p_S)$$

after reordering types if necessary.

Now

$$\overline{\alpha}_1 = {}^qD(v_1, \ldots, v_s) = \overline{\alpha}_2$$

and so $\overline{\alpha}_1 = \overline{\alpha}_2$. Also

$$\gamma_1 = M_{1-q}(\overline{\boldsymbol{P}_{\cdot 1}}, 1/\boldsymbol{p}) = M_{1-q}((v_1, \ldots, v_s), (1/p_1, \ldots, 1/p_s))$$

and similarly

$$\gamma_2 = M_{1-q}(\overline{\boldsymbol{P}_{\cdot 2}}, 1/\boldsymbol{p}) = M_{1-q}((v_1, \ldots, v_s), (1/kp_1, \ldots, 1/kp_s)) = \gamma_1/k.$$

Hence $\gamma_1 = k\gamma_2$, as claimed.

### S2.2.12 Sexual contact network

In Table S2.1(a), some of the inequalities (denoted by $\leq^*$) were only asserted to hold in the naïve-type model $\boldsymbol{Z} = \boldsymbol{I}$. Here we show that they all fail for general $\boldsymbol{Z}$. We do this using a single example.

Put $S = 6$, $N = 2$,

$$\boldsymbol{Z} = \begin{pmatrix} 1 & .5 & .5 & .7 & .7 & .7 \\ .5 & 1 & .5 & .7 & .7 & .7 \\ .5 & .5 & 1 & .7 & .7 & .7 \\ .7 & .7 & .7 & 1 & .5 & .5 \\ .7 & .7 & .7 & .5 & 1 & .5 \\ .7 & .7 & .7 & .5 & .5 & 1 \end{pmatrix}, \qquad \boldsymbol{P} = \begin{pmatrix} 1/6 & 0 \\ 1/6 & 0 \\ 1/6 & 0 \\ 0 & 1/6 \\ 0 & 1/6 \\ 0 & 1/6 \end{pmatrix}.$$

Then for all $0 < q < \infty$ and $j \in \{1, 2\}$,

$$\alpha_j = A = 3, \qquad \overline{\alpha}_j = \overline{A} = 1.5,$$
$$\rho_j = R = 2.05, \quad \beta_j = B = 0.488, \quad \overline{\rho}_j = \overline{R} = 1.025, \quad \overline{\beta}_j = \overline{B} = 0.976,$$
$$\gamma_j = G = 1.463$$

(to 3 decimal places). So in this example, all nine of the starred inequalities in Table S2.1(a) fail. We cannot, therefore, drop the hypothesis $\boldsymbol{Z} = \boldsymbol{I}$ under which those inequalities were stated. Nor can it be replaced by the weaker hypothesis that the triangle inequality holds ($Z_{i_1 i_2} Z_{i_2 i_3} \leq Z_{i_1 i_3}$ for all $i_1, i_2, i_3$), since this holds for the matrix $\boldsymbol{Z}$ above.

The key feature of this example is that each type is less similar to the types present in the same subcommunity than to the types present in the other subcommunity. One situation where this might occur is as follows.

Consider the spread of an infectious disease within a community. Within a given time period, the disease has probability $Z_{ii'}$ of being transmitted (perhaps indirectly) from individual $i$ to individual $i'$. Interpreting 'types' as individuals and using the uniform distribution $\boldsymbol{p}$ (so that all individuals are given equal weight), the diversity ${}^qD^{\boldsymbol{Z}}(\boldsymbol{p})$ provides a measure of the robustness of the community to the spread of infection.

Now take a simple heterosexual-only model of a sexually transmitted infection, involving a 'metacommunity' of six individuals divided into two 'subcommunities', one consisting of three males and the other consisting of three females. Since the infection cannot be transmitted directly between individuals of the same sex, the similarity coefficients $Z_{ii'}$ ($i \neq i'$) are lower when $i$ and $i'$ belong to the same subcommunity than when they belong to different subcommunities.

## References

[1] G. Hardy, J. E. Littlewood, and G. Pólya. *Inequalities*. Cambridge University Press, Cambridge, second edition, 1952.

[2] L. Jost. Partitioning diversity into independent alpha and beta components. *Ecology*, 88(10):2427–2439, 2007.

[3] L. Jost. Independence of alpha and beta diversities. *Ecology*, 91:1969–1974, 2010.

[4] T. Leinster and C. A. Cobbold. Measuring diversity: the importance of species similarity. *Ecology*, 93:477–489, 2012.

[5] H. Tuomisto. A diversity of beta diversities: straightening up a concept gone awry. Part 1. Defining beta diversity as a function of alpha and gamma diversity. *Ecography*, 33:2–22, 2010.